\documentclass[aps,prl,reprint,showpacs,preprintnumbers,amsmath,amssymb,superscriptaddress]{revtex4-1}

\usepackage{graphicx}
\usepackage{dcolumn}
\usepackage{bm}
\usepackage{helvet}
\usepackage{subfiles}
\usepackage{hyperref}
\usepackage{xcolor}

\hypersetup{
    colorlinks=true,
    linkcolor={blue},
    filecolor={red},      
    urlcolor={blue},
    citecolor={red},
    pdftitle={Overleaf Example},
    pdfpagemode=UseNone
    }

\begin{document}

\title{Measuring Chemical Shifts with Energy-Dispersive X-ray Spectroscopy}

\author{Yueyun Chen}
\affiliation{Department of Physics and Astronomy, University of California, Los Angeles, California 90095, USA}
\affiliation{California NanoSystems Institute (CNSI), University of California, Los Angeles, California 90095, USA}

\author{Rebekah Jin}
\affiliation{Department of Physics and Astronomy, University of California, Los Angeles, California 90095, USA}

\author{Yarin Heffes}
\affiliation{Department of Physics and Astronomy, University of California, Los Angeles, California 90095, USA}

\author{Brian Zutter}
\affiliation{Department of Physics and Astronomy, University of California, Los Angeles, California 90095, USA}
\affiliation{California NanoSystems Institute (CNSI), University of California, Los Angeles, California 90095, USA}

\author{Tristan P. O'Neill}
\affiliation{Department of Physics and Astronomy, University of California, Los Angeles, California 90095, USA}
\affiliation{California NanoSystems Institute (CNSI), University of California, Los Angeles, California 90095, USA}

\author{Jared J. Lodico}
\affiliation{Department of Physics and Astronomy, University of California, Los Angeles, California 90095, USA}
\affiliation{California NanoSystems Institute (CNSI), University of California, Los Angeles, California 90095, USA}

\author{B. C. Regan}
\affiliation{Department of Physics and Astronomy, University of California, Los Angeles, California 90095, USA}
\affiliation{California NanoSystems Institute (CNSI), University of California, Los Angeles, California 90095, USA}

\author{Matthew Mecklenburg}\email{mmecklenburg@cnsi.ucla.edu}
\affiliation{California NanoSystems Institute (CNSI), University of California, Los Angeles, California 90095, USA}

\date{\today}

\begin{abstract}
Electron microscopy prevalently uses energy-dispersive x-ray spectroscopy (EDS) and electron energy loss spectroscopy (EELS) for elemental analysis. EDS and EELS energy resolutions are commonly between 30-100 eV or 0.01-1 eV, respectively. Large solid angle EDS detector technology has increased collection efficiency to enable precision spectroscopy via averaging of 0.02-0.1 eV. This improved precision gives access to chemical shifts; examples are shown in compounds of Al, Ti, and W. EDS can now detect chemical information in a complementary parameter space (accelerating voltage, thickness, atomic number) to that covered by EELS.
\end{abstract}
\maketitle

It is not enough to know what elements are present, their chemical bonding is also important. Transmission electron microscopy (TEM) can resolve atomic-level defects and interfaces inaccessible to other microscopy methods. These boundaries contain important chemical information \cite{muller_mapping_1993}. X-ray spectroscopies for chemical detection, with energy resolution better than an eV, use absorption, emission, or photoelectrons and are spatially limited by x-ray sources and optics to micrometer probe sizes or greater \cite{sanchez-lopez_characterisation_1998,braun_origin_2002,lindgren_chemical_2004}. Spatial resolution using x-ray sources often does not surpass diffraction-limited optical microscope resolution \cite{blomfield_spatially_2005}. In the far field, only TEM can access elemental and chemical information at the atomic level.

Scanning transmission electron microscopy (STEM) rasters a sub-\r{a}ngstr\"{o}m sized probe to generate STEM images in parallel with techniques like energy-dispersive x-ray spectroscopy (EDS) and electron energy loss spectroscopy (EELS) \cite{muller_atomic-scale_2008, goodge_sub-angstrom_2020, zaluzec_quantitative_2022, zaluzec_x-ray_2023}. This pervasive pair of complementary spectroscopies measures the same inelastic events from different perspectives. EELS measures core electrons' binding energies after a primary electron inelastically scatters from a quantum bound state into a magnetic prism that disperses electrons onto a camera. Combining high spatial and energy resolution delivers chemical shift mapping across atomically sharp interfaces \cite{muller_mapping_1993,colliex_atomic-scale_2010}. EDS collects the emitted photon when an electron from a higher shell fills the hole made vacant by ejection from the bound state via inelastic scattering. EDS has comparatively poor energy resolution to EELS, but can detect a wider variety of elements. 

\begin{figure}[h]
\includegraphics[width=\linewidth]{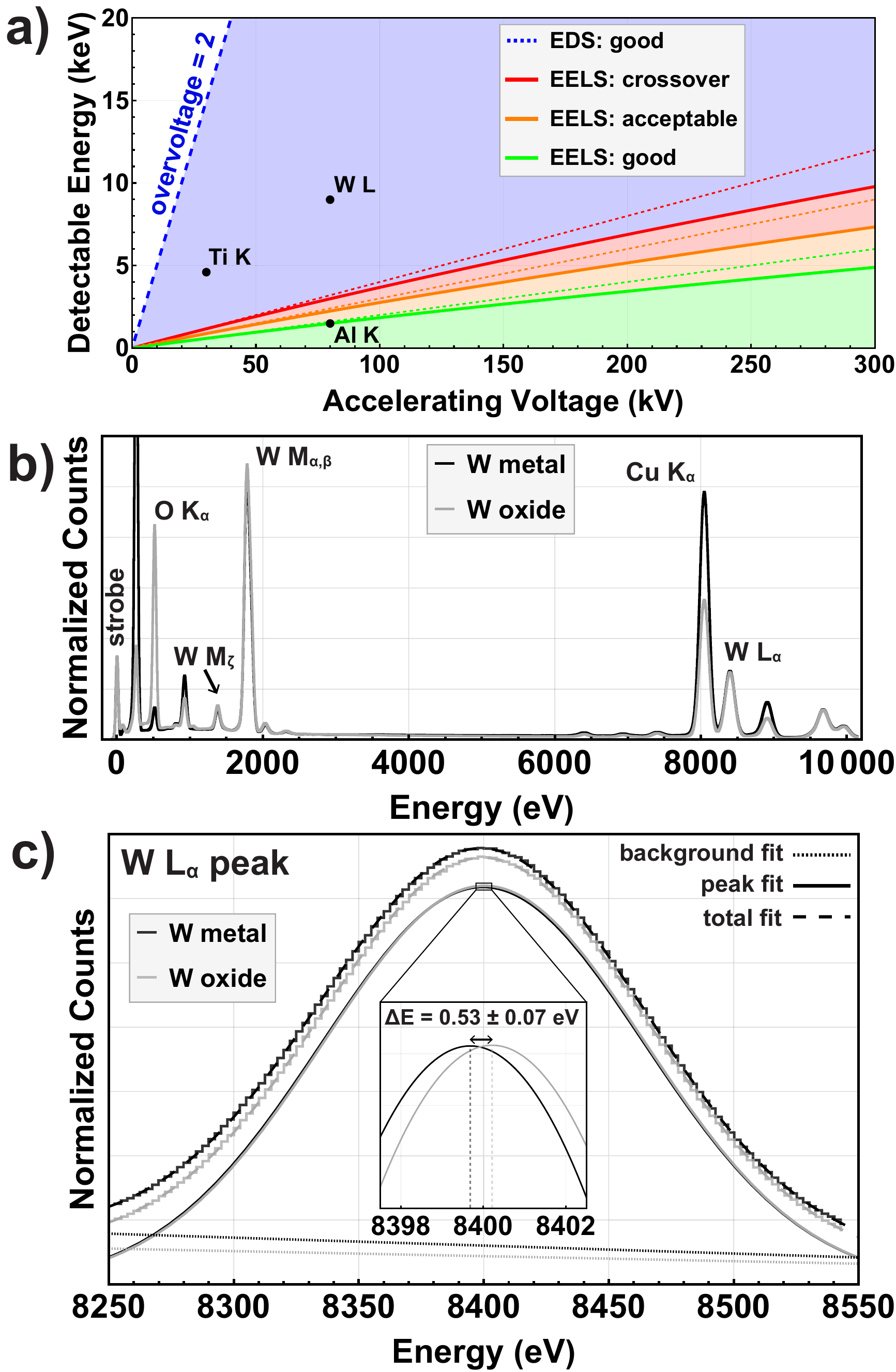}
    \caption{a) EDS and EELS detectable energy range plotted against the electron's accelerating voltage. EDS peak-to-background ratios and ionization cross sections are good with an overvoltage between 3-5 \cite{kohl_electronspecimen_2008}; 2 is shown as a lower bound. Approximate EELS good (10 mrads), acceptable (15 mrads) and crossover (20 mrads) cases for efficient collection angles are shown for relativistic (solid) and non-relativistic (dashed) cases \cite{maclaren_eels_2018}. b) EDS spectra from tungsten metal (black) and tungsten oxide (gray) acquired at 80 kV. c) Curve fitting (Gaussian and linear background) to the W L$_{\alpha}$ peak. A chemical shift of 0.53$\pm$0.07 eV is detected, and this shift is plotted inset.}
    \label{fig:eds}
\end{figure}

Electrons captured in EELS must pass through the sample. To keep the signal-to-background ratio (SBR) high the electrons should minimally scatter \cite{egerton_limits_2007, egerton_electron_2011}. Sharp edges (e.g. scattering from s and p orbitals) are not commonly accessible at binding energies with large quantum numbers and high scattering angles. High scattering angles can be defined as $\theta_s=\Delta E/2\gamma KE$, where $\Delta E/KE =$ 2\%, 3\%, and 4\% for good, acceptable, and threshold ranges, respectively \cite{maclaren_eels_2018} (Figure \ref{fig:eds}a, \footnote{$\gamma$ = 1+KE/m$_e$c$^2$, KE is the electron's kinetic energy, m$_e$ is the electron mass, and c is the speed of light}). What is inaccessible to EELS can be complementarily captured by EDS. Samples thick and thin can both produce good x-ray spectroscopy, only limited by the overvoltage needed to excite the transition (Figure \ref{fig:eds}a). X-ray spectroscopy's limited energy resolution has restricted it to elemental analysis. EDS chemical detection using electron sources, like that done using x-ray sources \cite{lindgren_chemical_2004}, can extend precision TEM spectroscopy into a new parameter space of thick samples, lower accelerating voltages, and transitions inaccessible to EELS.

Different x-ray spectrometer technologies have been used in TEM. Wavelength-dispersive x-ray spectroscopy (WDS) and microcalorimeters can both achieve sub-10 eV energy resolution \cite{castaing_application_1951,wollman_low_2001}. WDS circumscribes target, crystal analyzer, and detector along a Rowland circle. Through mechanical motion the analyser and detector move along the circumference to cover a broad range of energies. The limited solid angle, serial collection, and moving parts make WDS ineffective for TEMs. Microcalorimeters have cryogenic requirements and speed limitations making them (currently) unsuitable \cite{wollman_high-resolution_1998}. Silicon drift detectors (SDD) \cite{lechner_silicon_1996} replaced silicon-lithium detectors \cite{fitzgerald_solid-state_1968} due to their high speed, larger solid angles, and lack of cryogen requirements \footnote{See Supplimentary Info (SI) section Solid State Detector Evolution.}. SDD based EDS detectors are widely used on TEMs across the world and make up a large portion of the semiconductor industry's TEM workflow.

EDS instrumentation is less expensive than EELS by roughly an order of magnitude and requires no electron optics or associated alignments, making it more accessible than nearly any other spectroscopy \cite{zaluzec_x-ray_2023}. EDS's high SBR, even with thick samples or low accelerating voltages, makes it suitable for both scanning electron microscopes (SEMs) and TEMs \cite{goldstein_quantitative_1986}. At 30 kV accelerating voltage, accessible to SEMs and TEMs, EDS can already access the L line of all naturally occurring elements, whereas EELS efficiency (as in Fig. \ref{fig:eds}a) is restricted to elements below roughly titanium in atomic number. 

The SDD EDS \cite{lechner_silicon_1996} x-ray peak full width at half maximum (FWHM) is predominately determined by the electronic noise and shot noise fluctuation when converting photons into electron-hole pairs. An expression for this peak width measure is,
\begin{equation}
    FWHM_{EDS} = \sqrt{\left(2\sqrt{2 \ln{2} F \varepsilon_{x} \varepsilon_{i}}\right)^2+\varepsilon_{S}^2+\varepsilon_{LW}^2},
    \label{eq:FWHM}
\end{equation}
where $\varepsilon_{S} \approx$ 30 eV denotes the FWHM of the strobe peak (Fig. \ref{fig:eds}b), which arises from the electronic noise in the SDD \cite{goulding_aspects_1977}, $\varepsilon_{LW}$ denotes the natural line width of transition energy, which is typically a few eV and too small to matter in quadrature. The others parameterize the photon to electron-hole pair conversion. The Fano factor, $F \approx 0.12$, quantifies the ionization correlations in silicon \cite{shockley_problems_1961,klein_semiconductor_1968}. Parameters $\varepsilon_{x}$ and $\varepsilon_{i} (\approx$ 3.6 eV) are the x-ray energy and the silicon ionization energy, respectively \cite{lechner_ionization_1995,kotov_pair_2018}. 

For a W x-ray peak around 8 keV (W $L_{\alpha}$ for example, Fig. \ref{fig:eds}b), the EDS resolution $FWHM_{EDS} \approx$ 140 eV. This is two orders of magnitude larger than typical chemical shifts (a few eV or less) when compared linearly and much worse when added in quadrature. Insufficient resolution does not translate to insufficient precision. Since SDD are nearly shot noise limited, by averaging more x-ray events, the precision can be improved to a point where EDS becomes sensitive to sub-eV chemical shift. In Fig. \ref{fig:eds}c the measured chemical shift using curve fitting reveals a change just over half an eV  \footnote{See SI section Error Analysis of Energy for example fitting residuals}. Despite EDS being unable to resolve chemical shifts at the level of natural linewidth resolution, by collecting enough counts in the shot noise limited regime better than 100 meV energy precision is achieved for well-isolated peaks \footnote{The spectrometer bins and detector processing time are optimized to give the best precision and this process is discussed in the SI section Detector Optimization}. With enhanced energy precision, EDS can be used for chemical shift measurement on heavy elements and even chemical shift mapping complementary to EELS.

\begin{figure}[h]
\includegraphics[width=\linewidth]{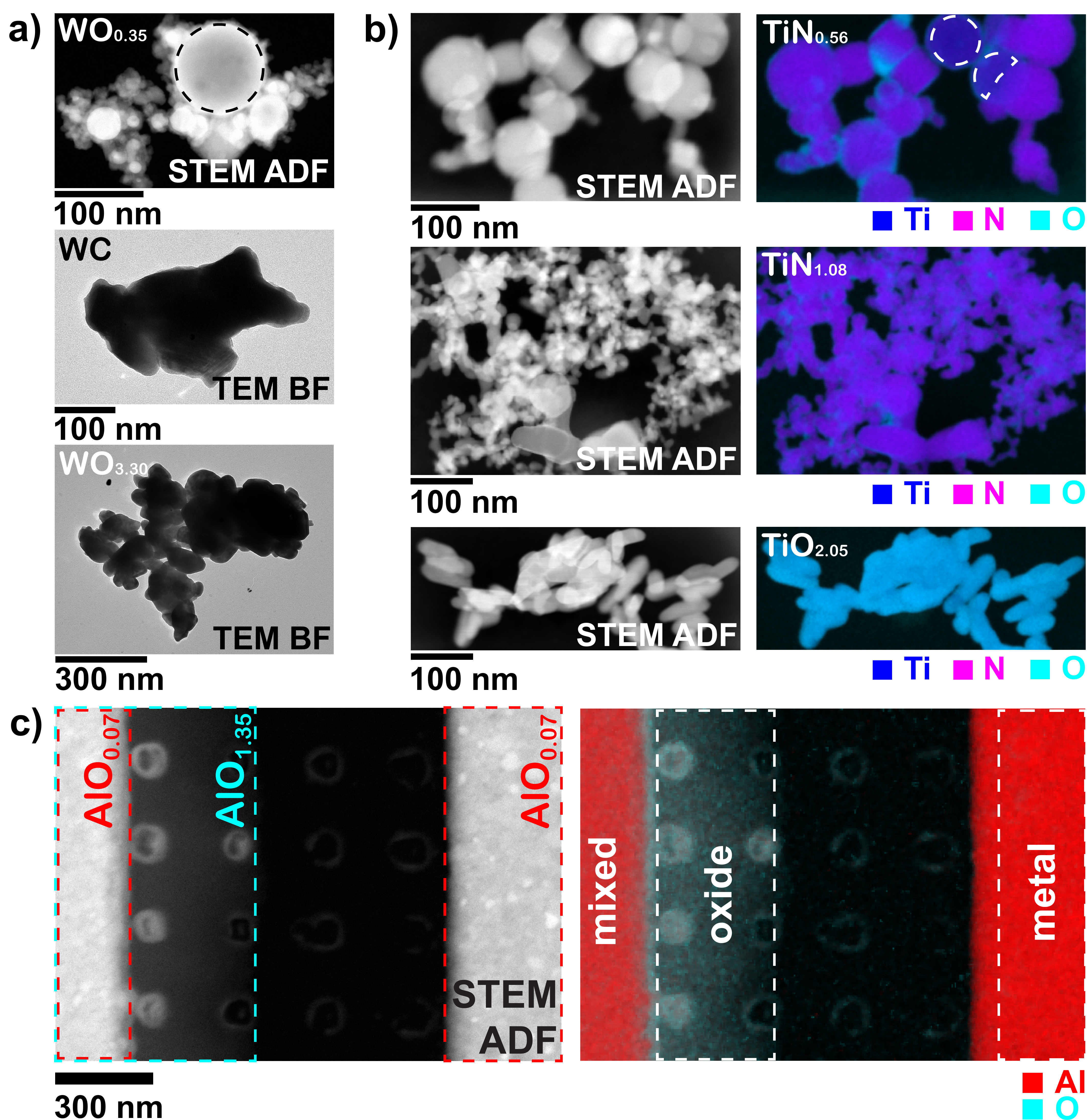}
  \caption{Bright field TEM or annular dark field STEM survey images and EDS element concentration maps of the tungsten, titanium, and aluminum specimens. The labelled chemical compositions are measured within the ROI or across the entire FOV when no ROI is present. a) Tungsten nanoparticles survey images. b) Titanium nanoparticles survey images and EDS maps. c) Aluminum films survey image and EDS map.}
  \label{fig:ptcl}
\end{figure}

To demonstrate EDS capability of measuring chemical shift, several compounds of tungsten (W), titanium (Ti), and aluminum (Al) are selected. The first two (acquired at at 80kV and 30kV, respectively) cover regions inaccessible with EELS. The three W compounds (Fig. \ref{fig:ptcl}a) and three Ti compounds (Fig. \ref{fig:ptcl}b) are commercially available nanoparticles deposited on transmission electron microscope (TEM) grids with a thin carbon film (\textless 10 nm) using standard drop-casting methods \footnote{See SI section on Sample Parameters for sample preparation and acquisition conditions.}. An aluminum film and an aluminum oxide film are deposited on a silicon chip with silicon nitride electron-transparent window using electron-beam evaporation and atomic layer deposition respectively (Fig. \ref{fig:ptcl}c). The Al sample is micro-fabricated and the patterns of the two Al films are defined by optical lithography \cite{hubbard_stem_2018}. Having multiple compounds of the same metal element allows us to see how the transition energy is affected by different chemical environments.

For each W specimen, a spectrum is acquired in the highlighted region of interest (ROI) or across the entire field of view (FOV). For an EDS spectrum like the one shown in Fig. \ref{fig:eds}b, the area under a certain peak is proportional to the abundance of the corresponding element, and any horizontal shift in the peak center is the chemical shift in the transition energy. For Fig. \ref{fig:ptcl}a top and bottom specimens, W L$_{\alpha}$ and O K$_{\alpha}$ peaks are used to determine the stoichiometry and measure the averaged valence of W. Following Cliff-Lorimer method \cite{cliff_quantitative_1975,goldstein_quantitative_1986} and assuming oxygen having its most common valence -2, the composition is determined to be WO$_{0.35}$ and WO$_{3.30}$ respectively, which corresponds to a W averaged valence of +0.7 and +6.6 respectively. However, the carbon support membrane makes valence measurement infeasible for tungsten carbide (Fig. \ref{fig:ptcl}a middle), so we take the valence to be +4 according to the manufacturer's 99.95$\%$ WC purity. For Ti and Al specimens, EDS spectrum images (where an EDS spectrum is acquired at each scan point in a raster pattern) are acquired, and elemental distributions are mapped. The valences of Al and Ti are determined the same way as the W specimens, and the chemical compositions are labelled in Fig. \ref{fig:ptcl}b,c \footnote{A comparison of valence values are tabulated in the SI section on Sample Parameters}.

\begin{figure}[h]
\includegraphics[width=\linewidth]{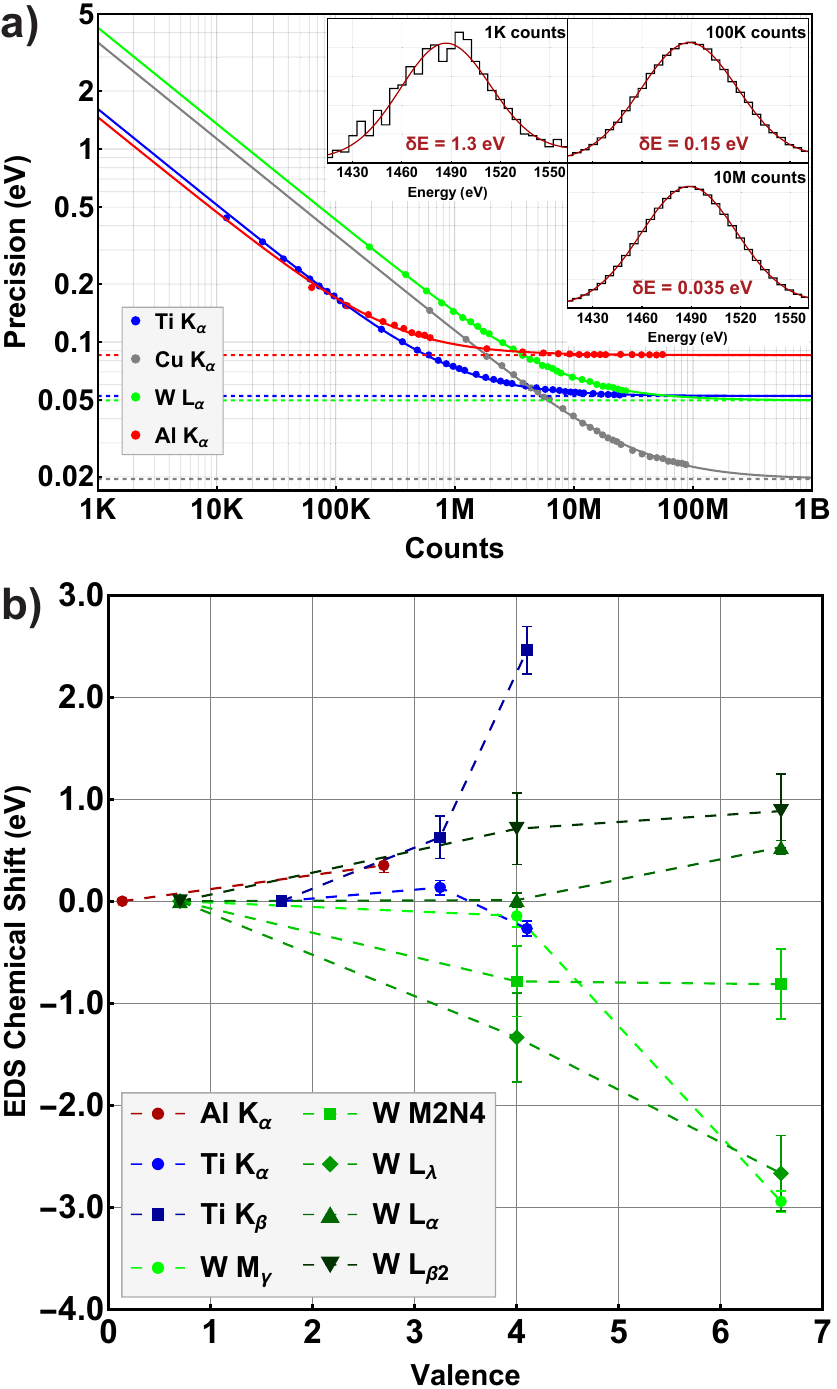}
  \caption{a) The precision (fit error in the mean) as a function of the number of x-ray counts detected for various elements and lines. The inset shows that as more x-ray events are collected, the Poisson noise is reduced, and the peak center can be determined more precisely. b) EDS chemical shifts as a function of valence measured between covalently bonded compounds shown in Fig. \ref{fig:ptcl}. EDS measures the chemical shifts in transition energies, and these shifts can be either positive or negative, unlike the EELS chemical shifts being universally positive.}
  \label{fig:plot}
\end{figure}

The resolution of EDS is limited by the peak energy energy and electronic noise (Eq. \ref{eq:FWHM}). An isolated peak can be measured more precisely than the FWHM just as the error in the mean decreases as die rolls are averaged. As the counts increase, curve fitting improves the precision similarly in a nonlinear way. This enhanced precision, typically three orders of magnitude better than the FWHM resolution, provides chemical shift detection in EDS. In the shot noise limited regime, the uncertainty carried by individual x-ray measurement disturbs the curve fitting and limits the precision. As the number of counts N increases, the precision goes up as a function of $\sqrt{N}$ (Fig. \ref{fig:plot}a). This relationship is governed by Poisson statistics. As N increases, the precision vs counts curve flattens and approaches a limit. This asymptotic behavior arises from the signal-to-background ratio (SBR) which is independent of the number of counts collected.

Solid state EDS detectors measure x-ray energy by collecting induced charges. A derived expression, the energy precision as a function of counts, is shown in Eq. \ref{eq:precision}. Poisson noise $\sigma_p$ from the counting process and precision limit $\sigma_0$ contribute. The uncertainty $\sigma_p$ comes from a single x-ray event. It is closely related to Equation (\ref{eq:FWHM}) empirically by $\sigma_p \approx 2 FWHM/(2\sqrt{2Ln2})$ \footnote{See section Detector Optimization is the SI for further analysis.}. Because of fluctuations in the ionization process, photons with the same energy do not always generate the same number of charges \cite{shockley_problems_1961}. This leads to the uncertainty $\sigma_p$ which is x-ray energy dependent, and its contribution to the total uncertainty can be reduced by collecting more x-ray events \cite{kallithrakas-kontos_x-ray_1996,xiao_possibility_1998}. In the expression is described how precision changes as a function of counts,
\begin{equation}
    \sigma_{precision} = \sqrt{\left(\frac{\sigma_p}{\sqrt{N}}\right)^2 + \sigma_0^2},
    \label{eq:precision}
\end{equation}
$\sigma_0$ represents highest precision can be achieved for a given peak, which is found to be negatively correlated with the SBR \footnote{see SI section Detector Optimization for SBR details.}. These two uncorrelated terms, $\sigma_p$ and $\sigma_0$, add in quadrature and determine the total uncertainty of the measured x-ray energy. This empirical expression fits multiple datasets (Fig. \ref{fig:plot}a). X-ray events are rare (one x-ray photon is generated per ten thousand beam electrons \cite{cosslett_radiation_1978}), but in solid state detectors collecting millions of counts is routine. 

Unlike EELS, where chemical shifts in the binding energies are always positive, EDS chemical shift in the transition energies can be either positive or negative \cite{lindgren_chemical_2004,yao_x-ray_1967,burkhalter_chemical_1972,kawai_high_1994,braun_origin_2002,boydas_chemical_2015,miaja-avila_valence--core_2021}. In three W samples imaged at 80 kV, with 2.5 eV/channel bin size, chemical shifts are observed in five transitions lines ranging from 1 to 10 keV (Fig. \ref{fig:plot}b). The three W L lines all have an energy higher than 7 keV, which is above the crossover regime for EELS (Fig. \ref{fig:eds}a). W L$_\alpha$ and W L$_{\beta 2}$ have positive chemical shifts less than +1 eV (about 1/3 of a spectrometer bin), and W L$_\lambda$ has a negative chemical shift down to -2.6 eV (about one bin). Chemical shifts are also observed in W M$_\gamma$ and W M2N4 lines, both negatively signed. Because of the high elastic/inelastic scattering cross section ratio and a short mean free path, W edges are very difficult to measure precisely with EELS. As an atom forms more bonds, the energies of core electrons are further minimized, which results in higher binding energies \cite{carpenter_structural_1982}. So, EELS chemical shifts are always positive \cite{sapre_chemical_1972}. While EDS measures the energy difference between two electron orbits, whether the chemical shift is positive or negative depends on which electron’s energy is minimized more when forming a new chemical bond. If the energy of the electron with a lower orbit is minimized more, the EDS chemical shift is positive, and vice versa. Thus, the observation of both positive and negative chemical shift coincides with our expectation.

In spectrum images acquired on Ti samples, spectra are summed inside the ROI or across the entire FOV. For Fig. \ref{fig:ptcl}b top, a ROI is applied to select nanoparticles with low nitrogen concentration to differentiate from Fig. \ref{fig:ptcl}b middle sample. The peak energies of Ti K$_\alpha$ and Ti K$_\beta$ lines are measure by curve fitting. The Ti K$_\beta$ line shifts up to +2.4 eV (about one bin), while the Ti K$_\alpha$ shifts less than 0.5 eV (Fig. \ref{fig:plot}). Acquired at 30 kV accelerating voltage, the 5 keV lineline is well over the crossover regime for EELS (Fig. \ref{fig:eds}a), but EDS can still provide precise information about the K shell. W and Ti measurements in the parameter space of large transition line and low accelerating voltage are inaccessible with EELS. By measuring Al K$_\alpha$ line around 1.5 keV (which is accessible with other TEM methods), the results are comparable with the literature. The spectrum image shows three distinct regions: mixed, oxide, and metal region (Fig. \ref{fig:ptcl}c). The two ROIs in Fig. \ref{fig:ptcl}c bottom highlight the regions where only aluminum or aluminum oxide film exists. The spectra in the two ROIs are summed respectively, and the Al K$_\alpha$ chemical shift is determined to be +0.35 eV by curve fit, which is consistent with previous measurements \cite{burkhalter_chemical_1972,lindgren_chemical_2004} (the shift is about 1/14 the bin size). A summary of the EDS chemical shifts is tabulated in the Supplementary Information. 

\begin{figure}[h]
    \includegraphics[width=\linewidth]{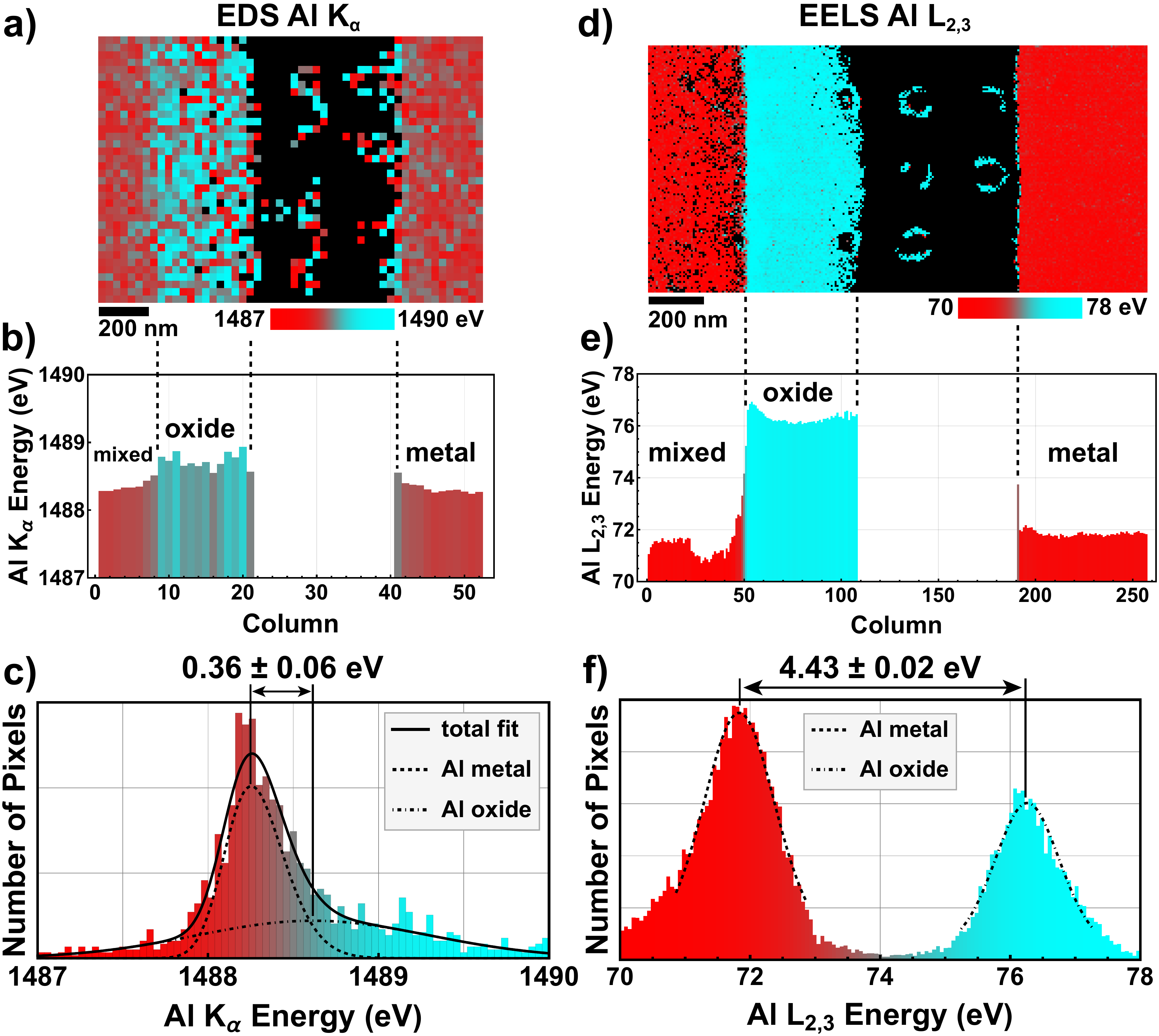}
    \caption{a) and d), EDS Al K$_\alpha$ energy map and EELS Al L$_{2,3}$ energy map. b) and e), column average plots of the energy maps above (columns with less than 50\% convergent fits were omitted). Three distinct regions are visible in both EDS and EELS plots. c) and f), histograms of the energy maps a) and d). Both histograms are fitted to two peaks representing the Al metal and oxide respectively. The EDS chemical shift is detectable despite being 12 times smaller than the EELS chemicals shift.}
    \label{fig:map}
\end{figure}

Finally, to demonstrate our capability to map chemical shift, the aluminum EDS spectrum image is binned to have enough counts to fit the Al K$_\alpha$ line on pixel level. Three distinct regions are visible in the Al K$_\alpha$ energy map (Fig. \ref{fig:map}a) together with the column-averaged plot (Fig. \ref{fig:map}b). The oxide region having a higher transition energy than the metal region indicates the Al K shell energy is minimized more than the L shell during the oxidation. A smooth transition is observed between the oxide and the mixed region and an oxide edge is detected at the edge of the metal region (Fig. \ref{fig:map}b). Even though the two peaks representing Al metal and oxide are not well-separated in the histogram (Fig. \ref{fig:map}c), a skewed normal distribution (for Al metal) and a normal distribution (for Al oxide) can be fitted to the distribution, and a shift of +0.36 eV is measured this way between the peak centers. This measurement on the histogram is in excellent agreement with the summed spectra measurement shown in Fig. \ref{fig:plot}b. And it demonstrates that chemical shift mapping is possible with a 0.6 Sr solid angle single crystal detector \footnote{Another example of mapping is shown in SI section EDS Chemical Shift Map.}. 

An EELS spectrum image is collected in the same region under the same microscope settings to compare with the EDS analysis. At 80 kV accelerating voltage, EELS is borderline ``acceptable" for measuring Al K edge at 1.5 keV, so only the Al L$_{2,3}$ edge (around 75 eV) is captured \footnote{An EELS chemical shift from the data in Fig. \ref{fig:map} is shown in Sup Fig. 10}. The Al L$_{2,3}$ edge, having a larger cross section, is much easier to map than the Al K edge. Because EELS is optically defined, the spectrum must be acquired quickly to avoid drift in the optical components. EDS can be mapped nearly indefinitely as spectra are not subjected to optical shifts. By measuring the Al L$_{2,3}$ edge location, a similar set of energy map and plots are generated (Fig. \ref{fig:map}d-f). Three distinct regions are visible in the EELS Al L$_{2,3}$ energy map and column-averaged plot, and the two peaks for Al metal and oxide are well-separated in the histogram. EELS appearing to differentiate the metal and the oxide better is a result of the chemical shift in Al L$_{2,3}$ edge being an order magnitude greater than that in the Al K$_\alpha$ transition. It does not imply EELS having a better energy precision. In fact, for the Al metal region, the EDS energy map has 3 times better precision. EDS Al K$_\alpha$ line measures the Al K-L$_{2,3}$ transition energy, and EELS Al L$_{2,3}$ edge measures the Al L$_{2,3}$ binding energy. Combining them, the Al K 1s binding energy is determined to be 1559.95 $\pm$ 0.03 eV for Al metal and 1564.88 $\pm$ 0.04 eV for Al oxide, similar to results using EELS \cite{sanchez-lopez_characterisation_1998}. Being able to match EELS’ energy precision and access a much higher energy range, EDS is a great complementary for chemical shift analysis purposes. And in this specific case, EDS provides information EELS missing about the Al K shell with comparable precision.

Chemical analysis has long been the domain of EELS \cite{egerton_limits_2007, egerton_electron_2011}. As the accelerating voltage is lowered, the range of energies accessible with EELS decreases monotonically. EDS, in contrast, does not suffer from these constraints. EDS can be used over a wide range of accelerating voltages, sample thicknesses, and atomic numbers. Moreover, an EELS spectrum is defined optically. The magnetic prism used to disperse electrons with different energies, although $\mu$-metal armored, is highly susceptible to changes in the local magnetic field \cite{egerton_limits_2007}. These changes, along with scan deflections, can shift the spectrum in energy space enough to cause a systematic effect that appears as a chemical shift. The zero-energy position in the EDS spectrum, not being optically defined, is nearly imperturbable, which allows adding spectra together from any time or position without any misalignment in energy \cite{lechner_silicon_1996}. Typically, EDS (EELS) is acquired over many frames (one frame) and a drift correction applied between frames (rows). Last but not least, an EDS system is more than an order of magnitude cheaper than conventional EELS systems. All these factors, versatility, stability, and cost efficiency, make it desirable to extend chemical analysis into the domain of EDS.

Through collecting large numbers of counts, EDS chemical shifts can be measured from a variety of compounds with sub-100 meV precision. And the EDS chemical shifts in transition energies can be either positive or negative, depending on the specific transition and chemical environment. Furthermore, chemical shift mapping across an Al metal-oxide interface is demonstrated (Fig. \ref{fig:map}a-c) with a common SDD. When used by itself, EDS can detect chemical shifts in transition lines with sub-100 meV precision. And when paired with EELS, EDS can extend chemical shift analysis in binding energies to one or two orders of magnitude higher energy, without requiring a higher accelerating voltage or any advanced electron optics. The spatial resolution of EDS chemical shift mapping, fundamentally limited by the sample dependent x-ray generation volume, ranges from atomic level (sub-\r{a}ngstr\"{o}m) to a few nanometers. Chemical shift maps in this work, limited by number of counts per unit area under the Al K$_\alpha$ peak, have a spatial resolution of about 30 nm. State-of-the-art SDDs can collect more than a million counts per second \cite{zaluzec_quantitative_2022}. With such a detector, the minimal number of counts needed for chemical shift detection (10k per peak) can be collected in 100 ms. EDS chemical shifts, in addition to being important for electron microscopy, can also be used where in other system using the same spectroscopy technique, such as x-ray florescent imaging. 

This work was supported by the Semiconductor Research Corporation (SRC), National Science Foundation (NSF) award DMR-1611036, and NSF STC award DMR-1548924 (STROBE). Data were collected at the Core Center of Excellence in Nano Imaging (University of Southern California), the Electron Imaging Center for Nanosystems (EICN) (RRID:SCR$\_$022900) at the University of California, Los Angeles’s California for NanoSystems Institute (CNSI), and the Thermo Fisher Scientific (TFS) Nanoport. Samples were fabricated in the Integrated Systems Nanofabrication Cleanroom (ISNC) at the California NanoSystems Institute (CNSI). We would like to thank Lee Casalena and Huikai Cheng for their help understanding and operating the TFS Spectra Ultra S/TEM with Ultra-X EDS.

\bibliographystyle{apsrev4-1}
\bibliography{20240626_eds_cs_bibtex_unico_wo_mod}

\end{document}


\title{Supplementary Information: Measuring Chemical Shifts with Energy-Dispersive X-ray Spectroscopy}

\author{Yueyun Chen}
\affiliation{Department of Physics and Astronomy, University of California, Los Angeles, Los Angeles, California 90095, USA}
\affiliation{California NanoSystems Institute (CNSI), University of California, Los Angeles, California 90095, USA}

\author{Rebekah Jin}
\affiliation{Department of Physics and Astronomy, University of California, Los Angeles, Los Angeles, California 90095, USA}

\author{Yarin Heffes}
\affiliation{Department of Physics and Astronomy, University of California, Los Angeles, Los Angeles, California 90095, USA}

\author{Brian Zutter}
\affiliation{Department of Physics and Astronomy, University of California, Los Angeles, Los Angeles, California 90095, USA}
\affiliation{California NanoSystems Institute (CNSI), University of California, Los Angeles, California 90095, USA}

\author{Tristan O'Neill}
\affiliation{Department of Physics and Astronomy, University of California, Los Angeles, Los Angeles, California 90095, USA}
\affiliation{California NanoSystems Institute (CNSI), University of California, Los Angeles, California 90095, USA}

\author{Jared Lodico}
\affiliation{Department of Physics and Astronomy, University of California, Los Angeles, Los Angeles, California 90095, USA}
\affiliation{California NanoSystems Institute (CNSI), University of California, Los Angeles, California 90095, USA}

\author{B. C. Regan}
\affiliation{Department of Physics and Astronomy, University of California, Los Angeles, Los Angeles, California 90095, USA}
\affiliation{California NanoSystems Institute (CNSI), University of California, Los Angeles, California 90095, USA}

\author{Matthew Mecklenburg}\email{mmecklenburg@cnsi.ucla.edu}
\affiliation{California NanoSystems Institute (CNSI), University of California, Los Angeles, California 90095, USA}

\maketitle

\tableofcontents

\newpage
\section{Solid State Detector Evolution}
\begin{figure}[!htb]
\vspace{\SIspace in}    
    \includegraphics[width=\SIscale\linewidth]{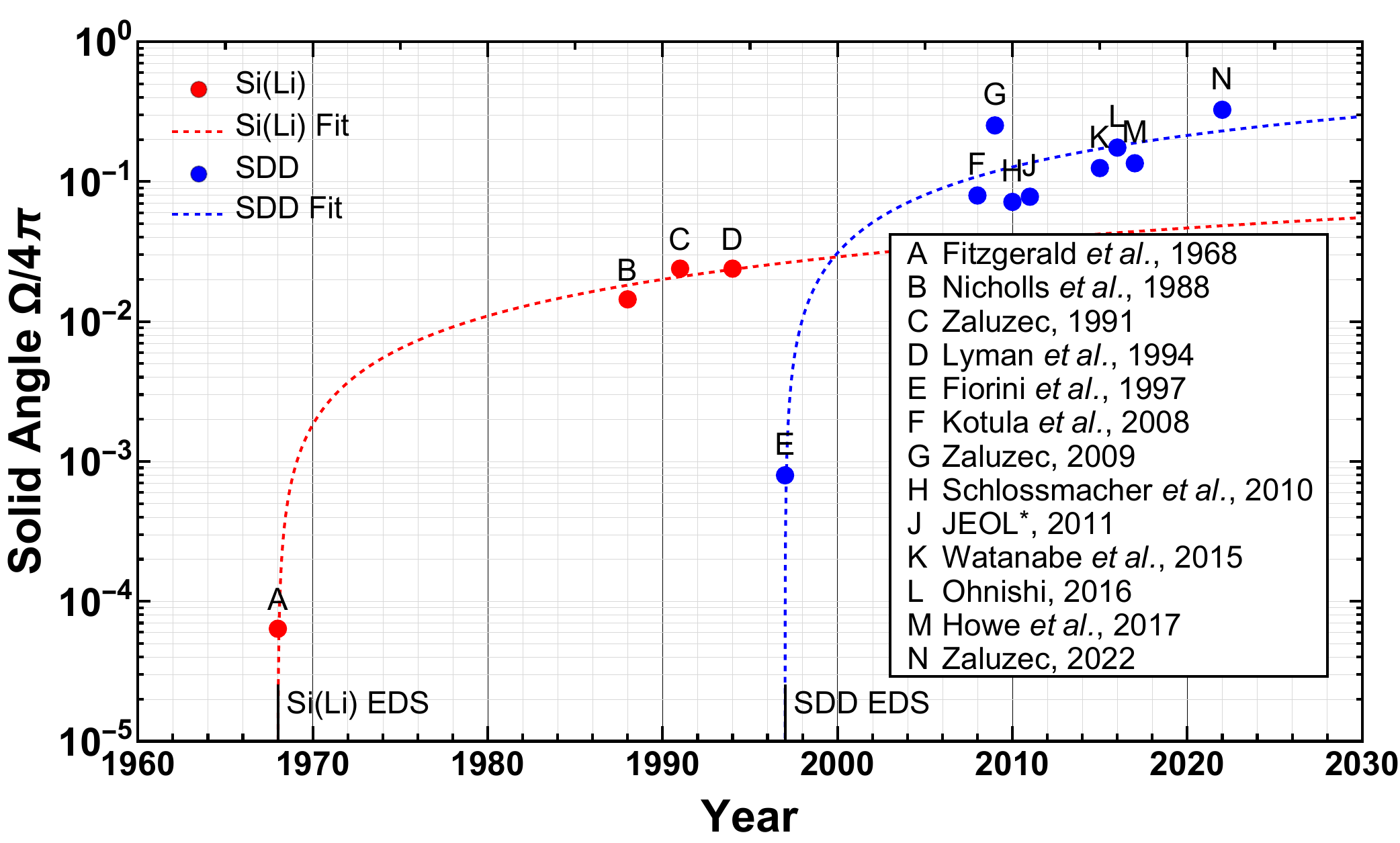}
    \captionsetup{width=\textwidth}
    \caption{The progress of x-ray detector's solid angle size over time, and a fit function for past interpolation and future extrapolate  \cite{fitzgerald_solid-state_1968,nicholls_comparison_1988,zaluzec_progress_1991,lyman_high-performance_1994,fiorini_new_1997,kotula_results_2008,zaluzec_innovative_2009,schlossmacher_nanoscale_2010,noauthor_product_2011,ohnishi_ultrahighly_2016,howe_collection_2017,zaluzec_quantitative_2022}. Silicon drift detectors (SDDs) have smaller collecting capacitors and larger detector areas. This gives access to faster acquisition speeds and larger solid angles than SiLi detectors.}
    \label{fig:detector}
\end{figure}

\newpage
\section{Error Analysis of Energy}

\begin{figure}[!htb]
\vspace{\SIspace in}    
    \includegraphics[width=\SIscale\linewidth]{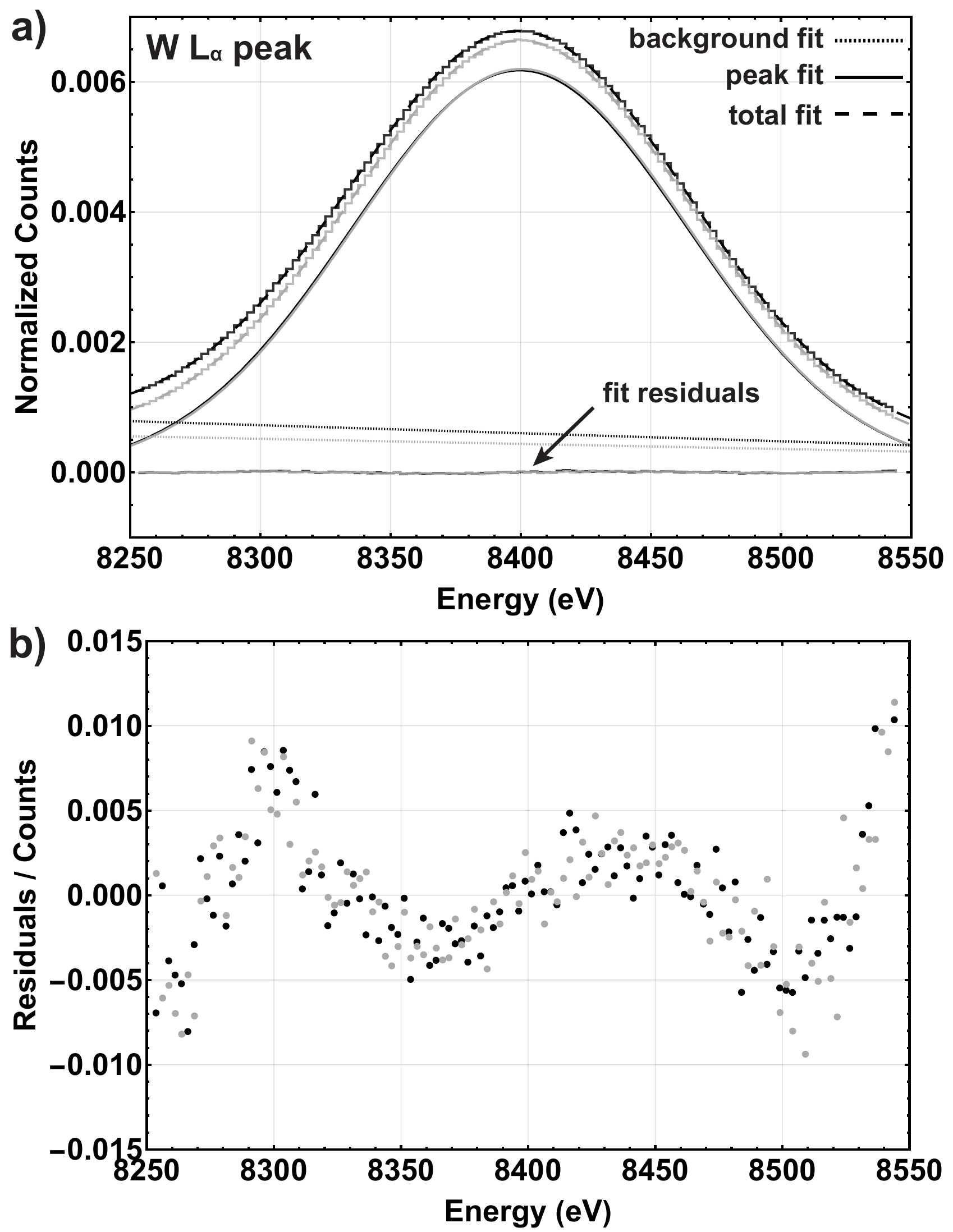}
    \captionsetup{width=\textwidth}
    \caption{Fit a) and residual b) from the data shown in main text Figure 1c. Black and gray represent W metal and oxide respectively. Normalized counts are used for the EDS data to allow comparison between metal and oxide. Unnormalized counts are used to calculate the residual to counts value shown in b).}
    \label{fig:Residuals}
\end{figure}

\begin{figure}[!htb]
\vspace{\SIspace in} 
    \includegraphics[width=\SIscale\linewidth]{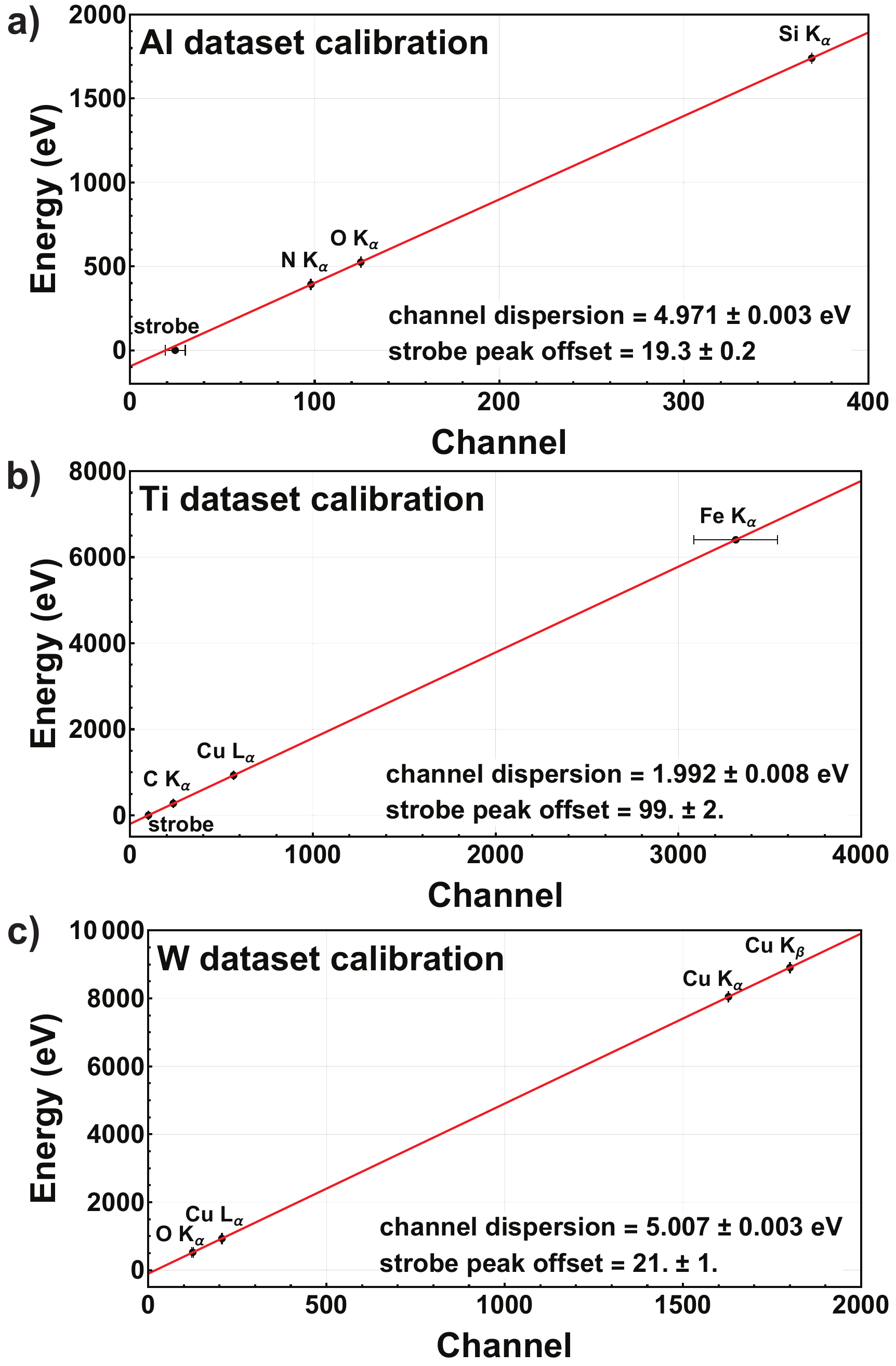}
    \captionsetup{width=\textwidth}
    \caption{Precise determination of the spectrometer channel size is necessary for accurately determining the x-ray energy. The bin size in each case is calibrated from background elements that were either uniformly distributed on the sample or from the microscope background. The calibrated values agree with manufacturer presets in each case (5 eV, 2 eV, 5 eV) to better than 1$\%$. Each spectrum is offset by the manufacturer to include the full dispersion of the strobe peak, which can be useful for determining the energy offset. However, the strobe peak is not always used to the energy offset because of detector specific background noise in the low energy regime. Background peaks that are closer to the peaks of interest are used for determining the offset for Al and W datasets.}
    \label{fig:CAL}
\end{figure}

Pulses from the charge sensing amplifier are sorted in a multi-channel analyzer. Each channel has an energy width set before data collection. The lowest energies are offset to enable detection of the strobe peak, which defines the zero energy. The x-ray energy is measured from the strobe to fluorescence peaks center to center. The spanned channels are multiplied by a calibration factor (in eV/ch, $D$) analogous to the dispersion in an electron energy-loss spectrometer. The fluorescent channels ($N_x$) are subtracted from the strobe channels ($N_s$) and multiplied by the calibration, $\Delta E_{accuracy} = D (N_x-N_s)$. The fraction error in each channel ($\delta D/D$) is smaller than 0.005 (based on the calibration fits in Figure \ref{fig:CAL}), however this parts-per-thousand bound limits the accuracy of our x-ray energy determination. Following error propagation in $\Delta E_{accuracy}$ ($D\delta N_x$ and $D\delta N_s$ are determined by the error in the center value) we derive,
\begin{equation}
\delta \Delta E_{accuracy} = D\sqrt{\left(\frac{\delta D}{D}\right)^2(N_x-N_s)^2 + \delta N_x^2 + \delta N_s^2},
\label{eq:fluc}
\end{equation}
and the dominating $\delta D/D$ term limits us to an eV or so because it is multiplied by a large span in channels $N_x-N_s$. 

But we are not attempting to determine the absolution energy of the x-rays. We are measuring the precision in a single peak and are comparing multiple measurements in that peak (here take as two for simplicity, A and B). As a result we only need to know the pixel values between the measurements, $\Delta E_{precision} = D (N_{x_A}-N_s) - D (N_{x_B}-N_s) = D (N_{x_A}-N_{x_B})$. The value of the strobe peak cancels as does its propagated error in 
\begin{equation}
\delta \Delta E_{precision} = D\sqrt{\left(\frac{\delta D}{D}\right)^2(N_{x_A}-N_{x_B})^2 + \delta N_{x_A}^2 + \delta N_{x_B}^2} \approx D \sqrt{\delta N_{x_A}^2 + \delta N_{x_B}^2} \approx \sqrt{2} D  \delta N_{x}.
\label{eq:flucAB}
\end{equation}
The previously limiting $\delta D/D$ term now is negligible in size because the shift's span in channels is small, typically less than one channel. The precision error in the shift is limited by the error from curve fit peak's center. Here we have calibrated the spectrum offset with a background peak with known energy for each dataset (Si K$_\alpha$ for Al, strobe peak for Ti, and Cu K$_\alpha$ for W). The offset peaks are selected based on SNR and proximity to the peaks of interest for higher measurement accuracy.

\begin{table}[!htb]
\vspace{\SIspace in} 
    \begin{tabular}{ |c |c |c |c |c| c| c| }
        \hline
        Element & Siegbahn & IUPAC (f-i) & Atomic (i-f) & X-Ray Energy [eV]  & Calc. FWHM [eV] \footnote{Using a 30 eV strobe peak} & Natural FWHM [eV] \\ [0.5ex] 
        \hline\hline
        Al & K$_{\alpha_1}$ & K-L$_{III}$ & 2p$^{3/2}$ $\rightarrow$ 1s$^{1/2}$ & 1486.708(10) \cite{deslattes_x-ray_2003} & 71.8 & 0.4 \cite{krause_natural_1979} \\
        \hline
        Al & K$_{\alpha_2}$ & K-L$_{II}$ & 2p$^{1/2}$ $\rightarrow$ 1s$^{1/2}$ & 1486.295(10) \cite{deslattes_x-ray_2003} & 71.8 & 0.4 \cite{krause_natural_1979}  \\
        \hline
        Ti & K$_{\alpha_1}$ & K-L$_{III}$ & 2p$^{3/2}$ $\rightarrow$ 1s$^{1/2}$ &  4510.8991(94) \cite{deslattes_x-ray_2003} &  108.6 & 1.2 \cite{krause_natural_1979}  \\
        \hline
        Ti & K$_{\alpha_2}$ & K-L$_{II}$ & 2p$^{1/2}$ $\rightarrow$ 1s$^{1/2}$ & 4504.9201(94) \cite{deslattes_x-ray_2003}  & 108.6 & 1.2 \cite{krause_natural_1979}  \\
        \hline
        Ti & K$_{\beta_{1,3}}$ & K-M$_{II,III}$ & 3p$^{1/2,3/2}$ $\rightarrow$ 1s$^{1/2}$ & 4931.827(59) \cite{deslattes_x-ray_2003}  & 113.2 & 2.9 \cite{roseberry_effects_1936} \\ 
        \hline
        W & M$_{\alpha_1}$ & M$_{V}$-N$_{VII}$   & 4f$^{7/2}$ $\rightarrow$ 3d$^{5/2}$ & 1775.5(3) \cite{bearden_x-ray_1967} & 83.8 & x \\
        \hline
        W & M$_{\alpha_2}$ & M$_{V}$-N$_{VI}$  & 4f$^{5/2}$ $\rightarrow$ 3d$^{5/2}$ & 1773.2(5) \cite{bearden_x-ray_1967} & 83.8 & x \\
        \hline
        W & M$_{\beta}$ & M$_{IV}$-N$_{VI}$   & 4f$^{5/2}$ $\rightarrow$ 3d$^{3/2}$ & 1834.9(3) \cite{bearden_x-ray_1967} & 84.7 & x \\ 
        \hline
        W & L$_{\alpha_1}$ & L$_{III}$-M$_{V}$  & 3d$^{5/2}$ $\rightarrow$ 2p$^{3/2}$ & 8398.242(54) \cite{bearden_x-ray_1967} & 151.1 & 7.5 \cite{akita_investigation_1995}\\
        \hline
        W & L$_{\alpha_2}$ & L$_{III}$-M$_{IV}$  & 3d$^{3/2}$ $\rightarrow$ 2p$^{3/2}$ & 8335.34(17) \cite{bearden_x-ray_1967} & 150.6 & 7.7 \cite{akita_investigation_1995}\\ 
        \hline
        W & M$_{\zeta_1}$ & M$_{V}$-N$_{III}$  &  4p$^{3/2}$ $\rightarrow$ 3d$^{5/2}$ & 1383.4(6) \cite{bearden_x-ray_1967} & 78.0 & x \\
        \hline
        W & M$_{\zeta_2}$ & M$_{IV}$-N$_{II}$  &  4p$^{1/2}$ $\rightarrow$ 3d$^{3/2}$ & 1378.7(8) \cite{bearden_x-ray_1967} & 78.0 & x  \\
        \hline
     \end{tabular}
    \caption{Notation differences in EDS transitions, the transition energies, the associated FWHM, and the corresponding natural linewidth. The x's indicate values that could not be found in the literature.}
    \label{fig:TRANS}
\end{table}

\section{Detector Optimization}
Due to the detection chain's finite reset time between x-ray strikes (called in physics parlance the dead time $\tau_D$), there is a maximum rate of x-ray detection ($1/\tau_D$) in these paralyzable pulse counters. The optimal counting dead time is $D \approx63 \%$ (derivation below). The probe size, sample thickness, specific transition, processing time, and beam current all determine the number of x-rays emitted per incident electron. This value is easiest set experimentally from the current which then controls the number of x-rays entering the detector chain and setting the counted rate $R_{out}$. The FWHM expression (Equation 1 in the main text) is at or near the ideal processing time set by the spectrometer settings.

\begin{figure}[!htb]
\vspace{\SIspace in}   
    \includegraphics[width=\SIscale\linewidth]{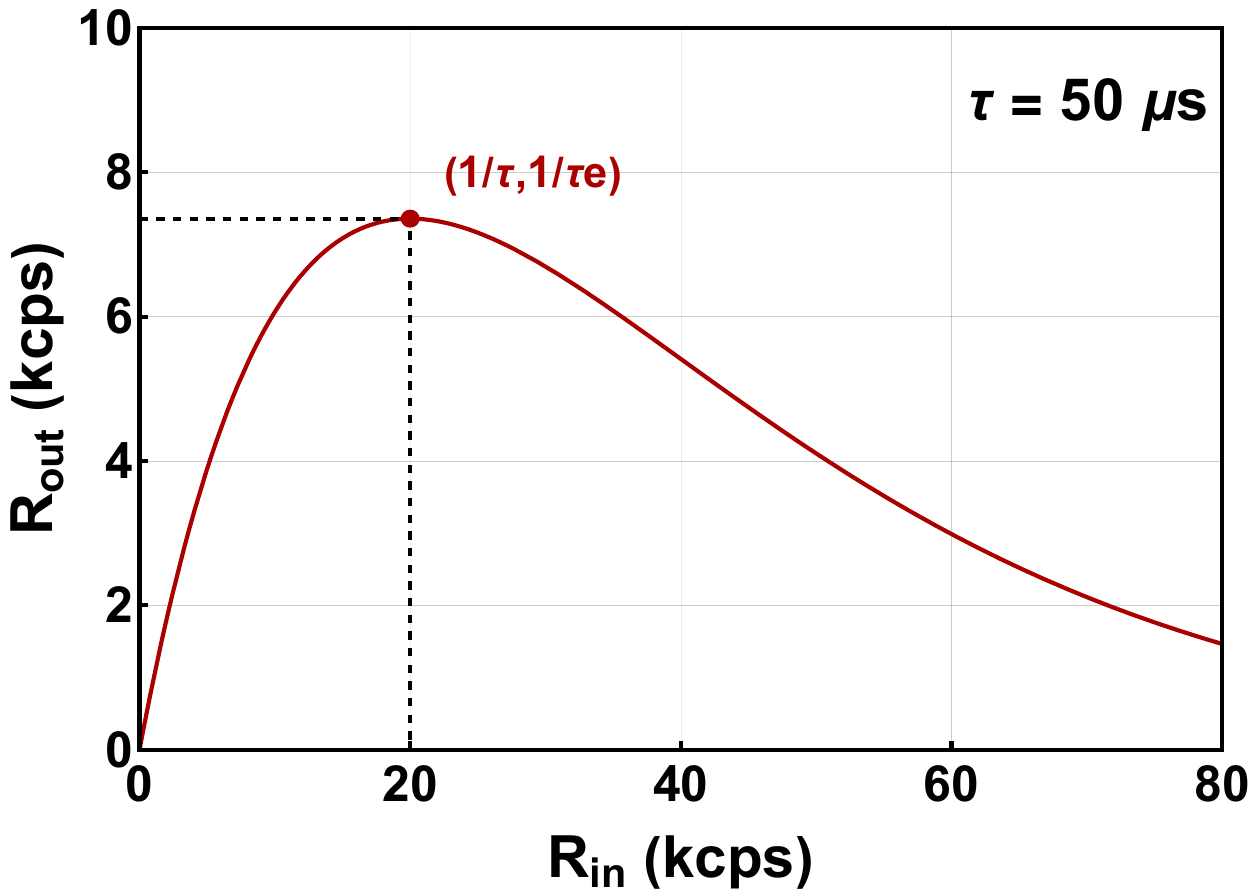}
    \captionsetup{width=\textwidth}
    \caption{Paralyzable detector output count rate as a function of the input count rate for the indicated dead time.}
    \label{fig:RINROUT}
\end{figure}

Pulse counting electronics are either non-paralyzable (the $R_{out}$ increases monotonically with $R_{in}$) or paralyzable ($R_{out}$ eventually decreases as $R_{in}$ increases). Assuming Poisson statistics, the incident flux into the detector is inoperable for a period $\tau_D$ after being hit by and x-ray. The relationship between $R_{in}$ and $R_{out}$ is (in the simplest approximation), 
\begin{equation}
R_{out} = R_{in}e^{-R_{in}\tau_D}.
\label{eqn:EDSrates}
\end{equation}
The dead time is not to be confused with the counting dead time $D$ (in microscopy parlance) showing how close the input and output count rates are matched,
\begin{equation}
D=\frac{R_{in}-R_{out}}{R_{in}}\times 100\%.
\end{equation}
The dead time $\tau_D$ can be controlled qualitatively by the `process time' parameter in the EDS software. This often dimensionless setting can effectively be set to reduce (increase) the dead time $\tau_D$ with a low (high) value. Decreasing the process time also monotonically decreases the width of the noise peak as shown in Supplementary Figure \ref{fig:PROCESS}.

Achieving the high precision measurements of the x-ray energies requires a large numbers of counts. To keep acquisition times reasonable, the count rate of measured x-ray events, $R_{out}$, is as high as possible, while the energy resolution remains optimal near the longest processing time. The different heights are shown in Figure \ref{fig:RINROUT}. The simple model in \ref{eqn:EDSrates} does not account well for lower process times (more input count rates, $R_{in}$). This is due to some subtleties in statistics that will be addressed in a following paper (the results are outside the scope here, the ideal case is close to our operating conditions). This input rate (for a given detector) is governed by the beam current, amount on inelastic scattering, and the x-ray's take off angle from sapmle to the detector. If $R_{in}$ is too high x-rays will begin to arrive so quickly that the pulse counting electronics will not be able to discriminate x-ray events, resulting in a decrease in $R_{out}$, in other words the detector's ability to count will be paralyzed.

For a given $\tau_D$ the maximum in Equation \ref{eqn:EDSrates} occurs when $R_{in}=\tau_D^{-1}$. However, the actual value of $\tau_D$ in seconds is often not provided by the EDS detector software but instead is left as proportional to a dimensionless parameter. The counting dead time $D$ at the optimal input rate is $D=(1-1/e)\times 100\%$. For any process time, $D_{optimal}\approx63\%$ corresponds to the maximum in the counting dead time and is the value used in this work.

\newpage
\begin{figure}[!htb]
\vspace{\SIspace in}
\includegraphics[width=\SIscale\linewidth]{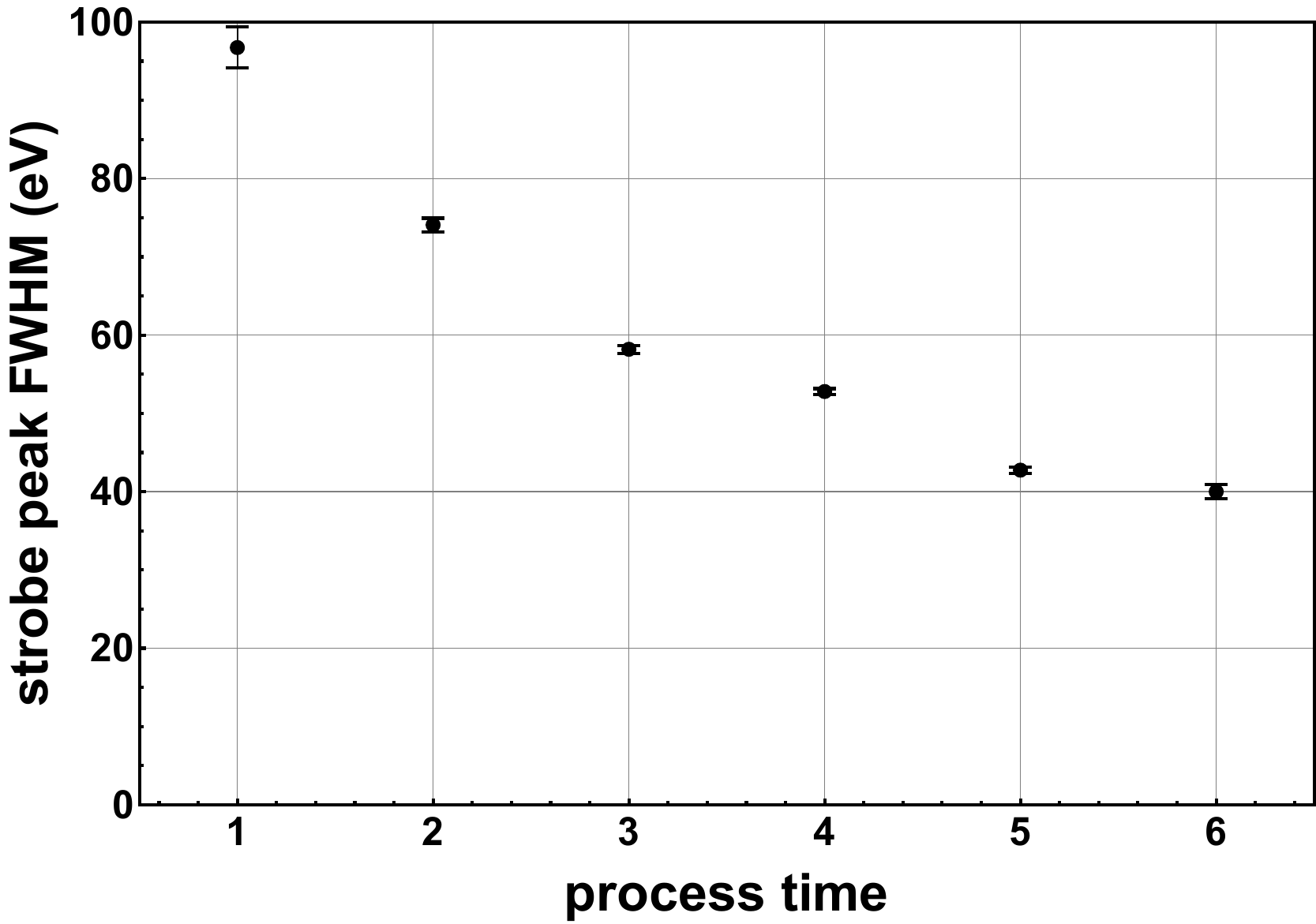}
    \captionsetup{width=\textwidth}
    \caption{The NiO$_x$ strobe peak as a function of the detector chains processing time. The processing time indicated here is a parameter that is proportional to the processing time used in Oxford's detector chain. Larger (longer) processing times produce a smaller noise peak monotonically. The longest processing times were used in nearly all acquisitions to ensure the best energy resolution.}
    \label{fig:PROCESS}
\end{figure}

\newpage
\begin{figure}[!htb]
\vspace{\SIspace in}
\includegraphics[width=\SIscale\linewidth]{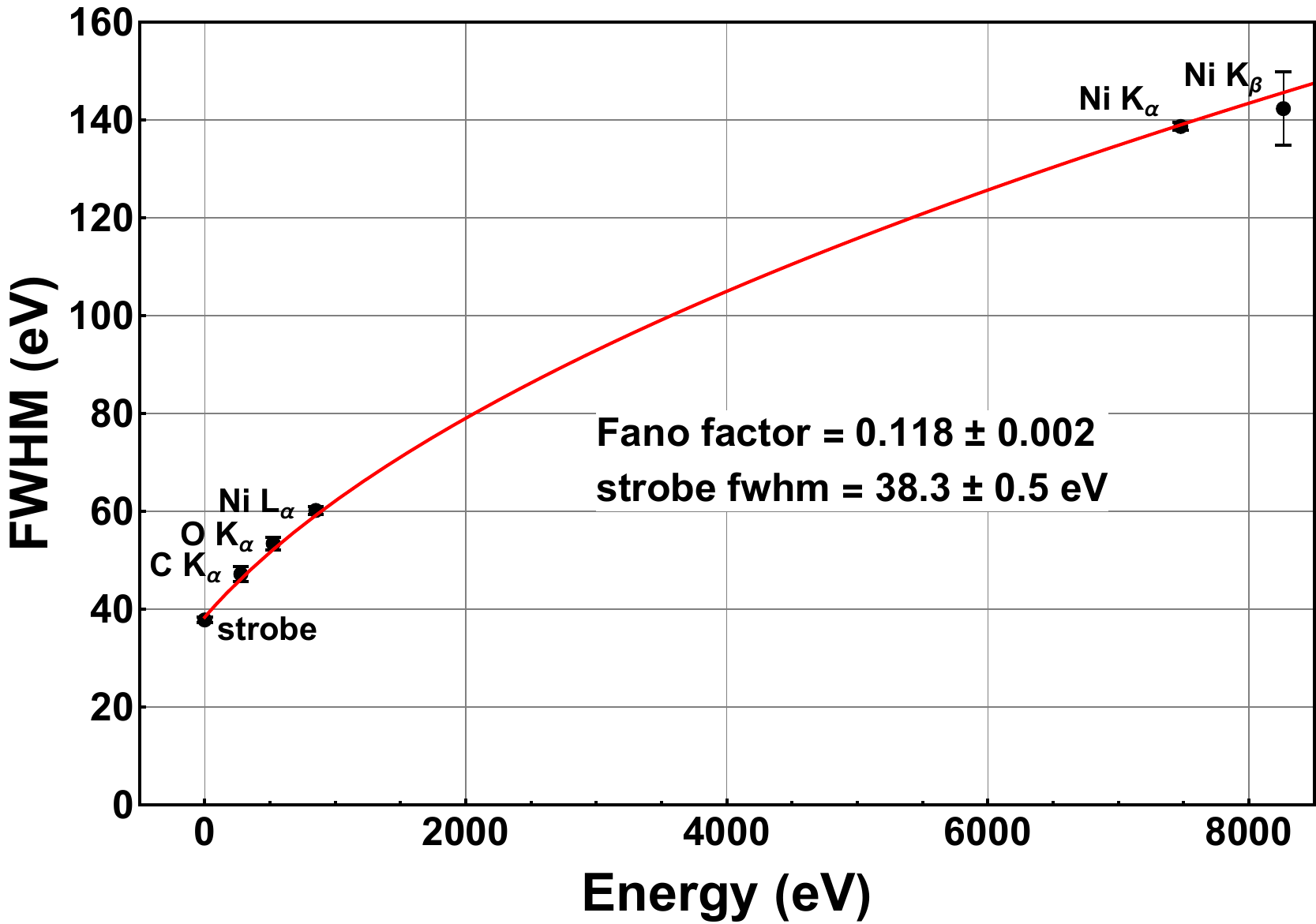}
    \captionsetup{width=\textwidth}
    \caption{The FWHM of the noise peak, C, O, Ni$_L$, and Ni$_K$ as a function of the measured energy of each peak. The FWHM and the center are measured by curve fitting to main text Equation 1. The fit parameters are the Fano factor and strobe peak FWHM, the Fano parameter compares well to known values for silicon \cite{mckay_electron_1953,shockley_problems_1961,mazziotta_electronhole_2008,ramanathan_ionization_2020}.}
    \label{fig:FWHM}
\end{figure}

\newpage

\newpage
\begin{figure}[!htb]
\vspace{\SIspace in} 
    \includegraphics[width=\SIscale\linewidth]{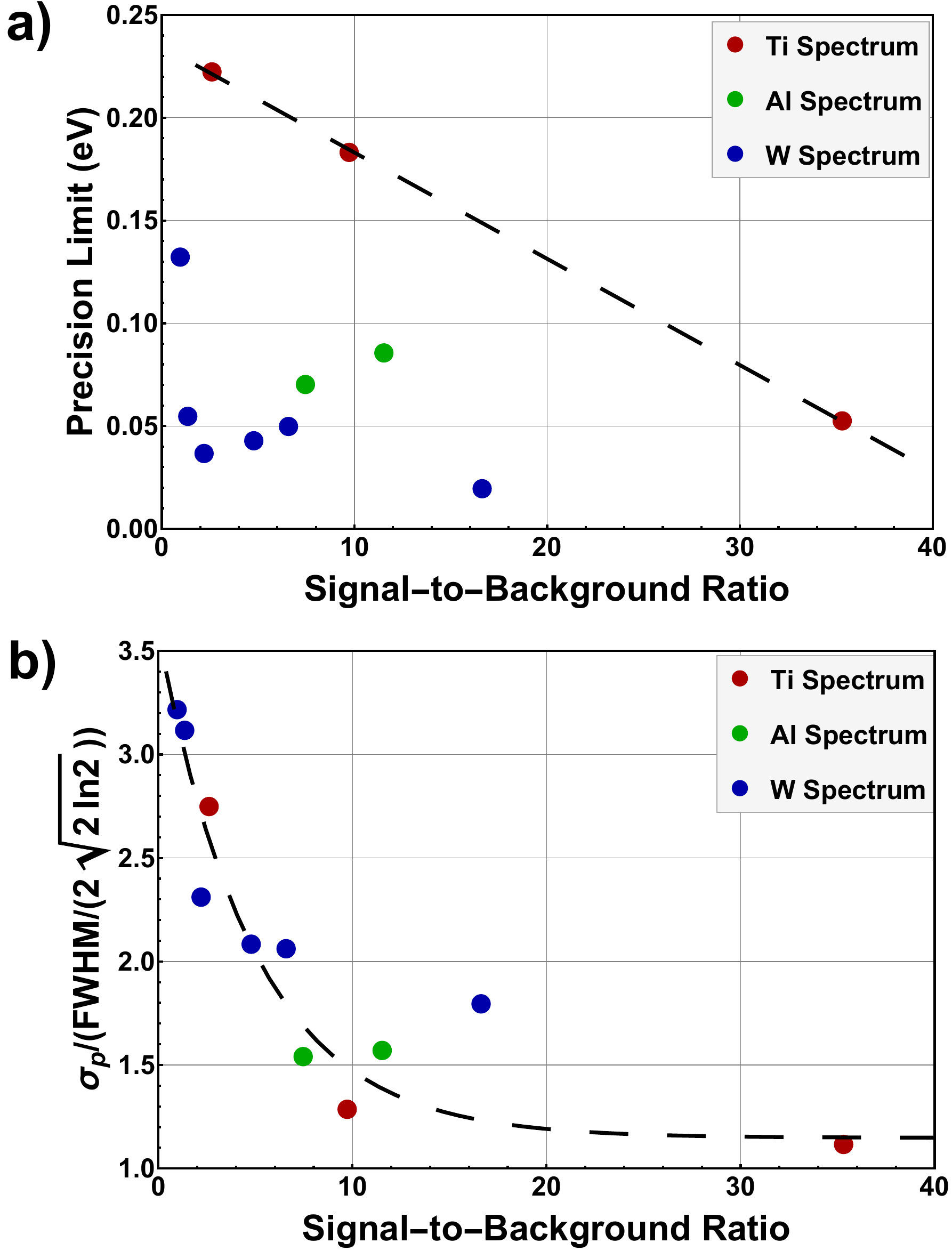}
    \captionsetup{width=\textwidth}
    \caption{The precision limit (in eV) and the normalized $\sigma_p$ peak standard deviation ratio are plotted as a function of the signal-to-background ratio (SBR). Semi-quantitative correlations are given by the dashed lines in the plot. Although the exact relationship between these quantities is not given, the correlation is clear: a better SBR leads to a better precision limit and better agreement with the peak standard deviation. The normalized $\sigma_p$ peak standard deviation ratio larger than 1 is affected by having a background model in the nonlinear fit, which leads to a larger error estimation for the signal model parameters. The standard deviation determined by the precision at one count, $\sigma_p$, is normalized to the standard deviation derived from the Gaussian FWHM.}
    \label{fig:SBR}
\end{figure}

\newpage
\section{Sample Parameters}
The two aluminum samples are built on silicon substrates. The silicon substrates are 200 $\mu$m thick with a 800 nm thick silicon dioxide film and a 20 nm thick silicon nitride (SiN) film on each side. SiN is the outermost layer on both sides. At the bottom side, the SiN is patterned with photolithograph and removed with reactive-ion etching (RIE). The SiN-removed region defines the electron transparent window needed for TEM imaging. KOH wet etch and HF vapor etch are used to remove the silicon and silicon dioxide in the SiN-removed region. Eventually, only the top SiN layer is left and becomes the membrane for the electron-transparent window. The patterns of both aluminum and aluminum oxide are defined with photolithography. 100 nm of aluminum is deposited using electron beam evaporation with CHA Mark 40 on the SiN membrane. And another 20 nm thick aluminum oxide film is put down using atomic layer deposition (ALD) with Ultratech Fiji Thermal and Plasma Atomic Layer Deposition System.

Tungsten and titanium samples are prepared with commercially available nanoparticles from SkySpring Nanomaterials, with detailed information shown in Table \ref{fig:SAMPLEPREP}. Each type of nanoparticles is deposit onto a 3 mm TEM grid with 20 nm-thick carbon membrane with drop-casting method. About 10 mg of nanoparticles are suspended in 1 mL of isopropyl alcohol, sonicated for 15 minutes, and centrifuged with Eppendorf Centrifuge 5417C. For each type of nanoparticles, 2 µL of suspension right above the precipitation is transferred to a TEM grid and dried. Considering the impurity introduced during manufacture and contact with air (especially for W and Ti metal nanoparticles), each type of nanoparticle is analyzed with hyperspectral EDS imaging to sort out the appropriate regions for chemical shift measurements.

\begin{table}[!htb]
    \begin{tabular}{ |c |c |c |c|  }
    \hline
    Product Number & Material & Diameter (nm) & Centrifuge \\ [0.5ex] 
    \hline\hline
    1123XH	&Ti	&40 – 60 &2500 RPM for 20 min \\
    \hline
    7945HK	&TiN &20 &1000 PRM for 1 min \\
    \hline
    7920DL	&TiO2 &10 – 30	&100 RPM for 10 sec \\
    \hline
    9821XH	&W &80 – 100 &800 RPM for 20 min \\
    \hline
    8005DX	&WC	&80 &800 RPM for 20 min \\
    \hline
    8010CN	&WO3 &$<$100	&500 RPM for 2 min \\
    \hline
  \end{tabular}
    \caption{Table of sample preparation parameters.}
    \label{fig:SAMPLEPREP}
\end{table}

\newpage
\begin{figure}[!htb]
\vspace{\SIspace in} 
    \includegraphics[width=\linewidth]{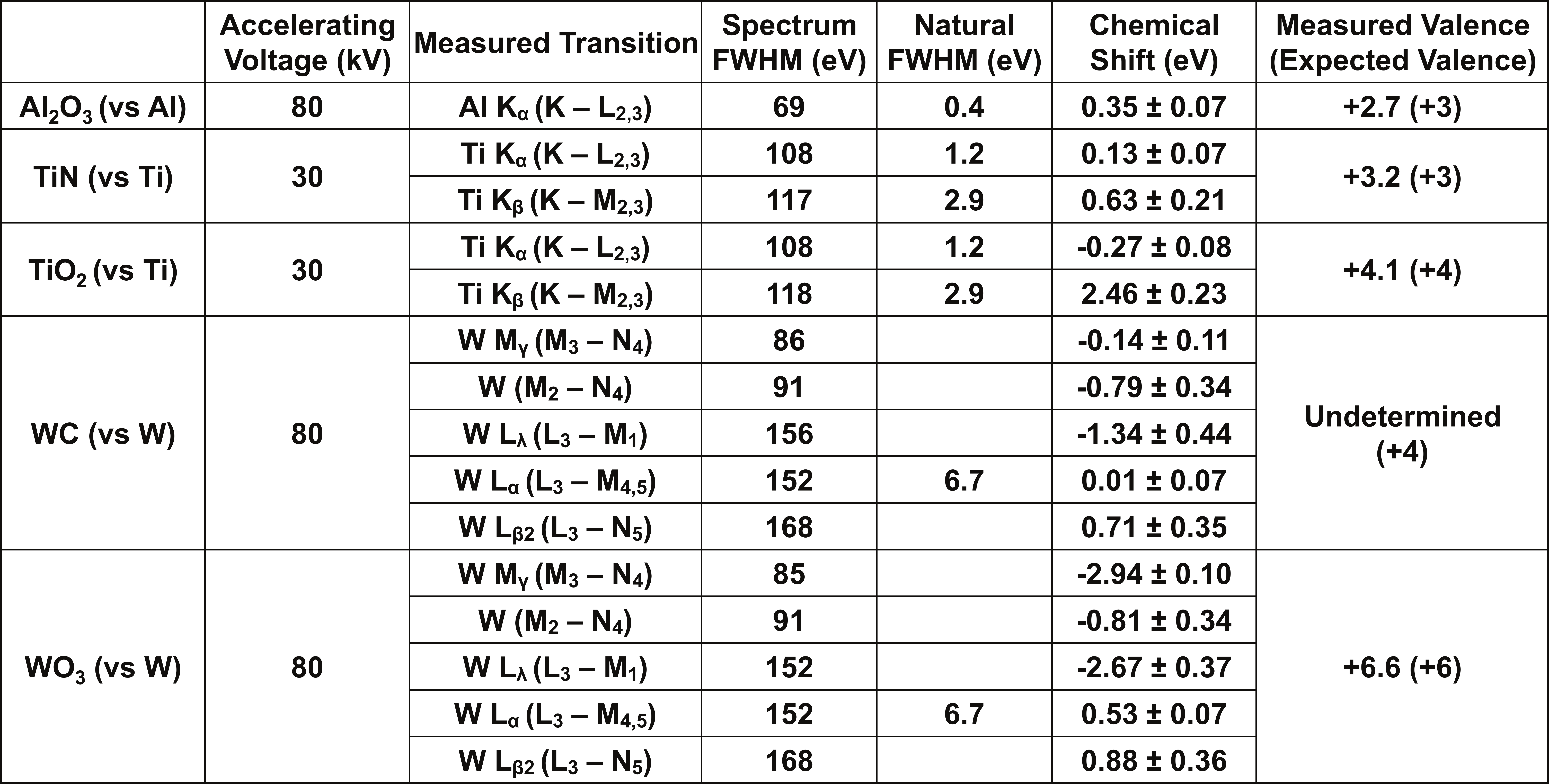}
    \captionsetup{width=\textwidth}
    \caption{A table comparing the dataset, accelerating voltage at which it was taken, the measured transition, the spectrometers FWHM, the natural linewidth FWHM, the measured chemical shift, and the measured (expected) valence. Al, and W data were collected on Oxford X-MaxN 100TLE 100 mm$^2$ SDD, the Ti data was collected on the TFS Ultra-X EDS.}
    \label{fig:summary}
\end{figure}

\newpage
\section{EDS Chemical Shift Map}
\begin{figure}[!htb]
\vspace{\SIspace in}    
    \includegraphics[width=\SIscale\linewidth]{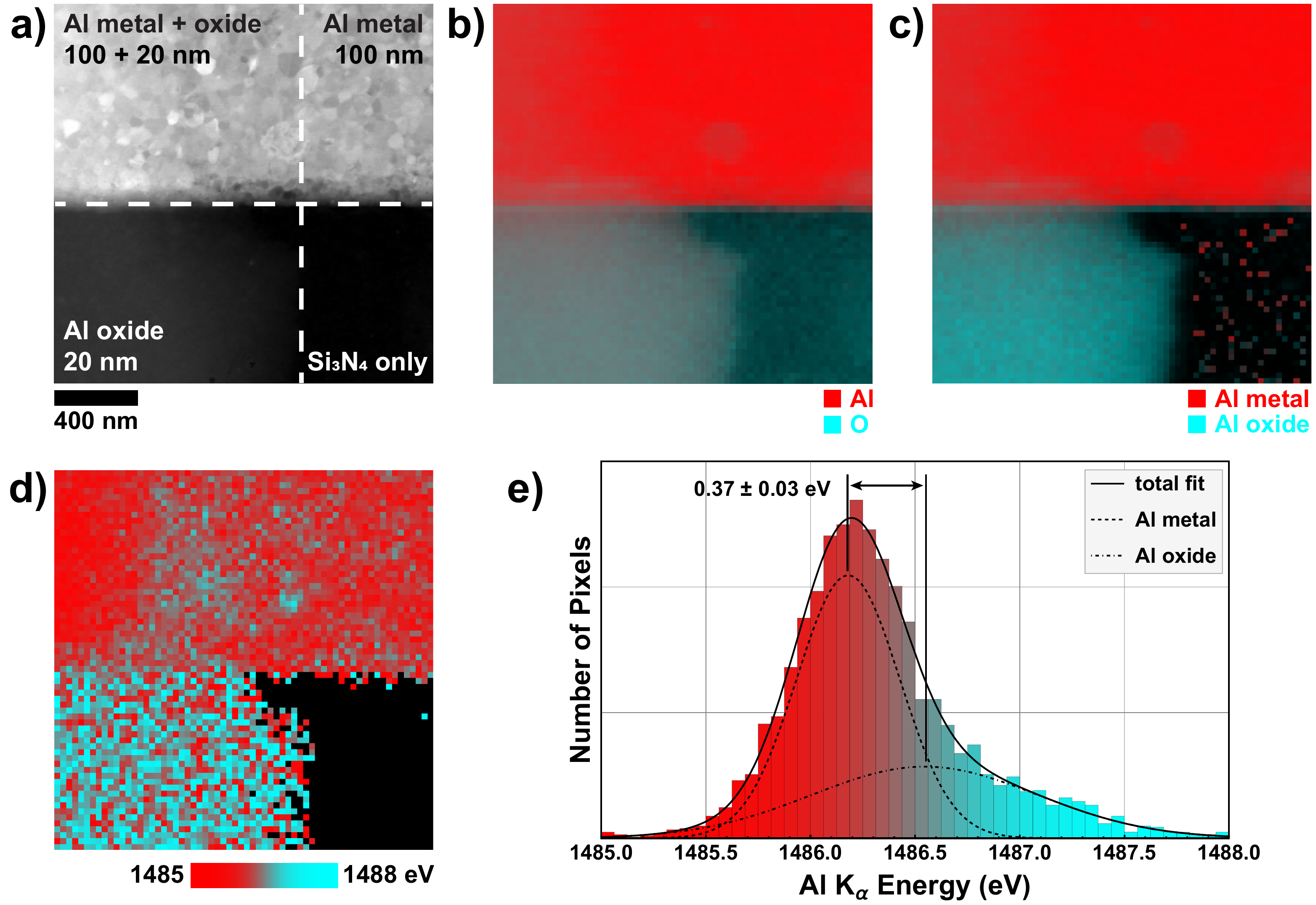}
    \captionsetup{width=\textwidth}
    \caption{A EDS chemical shift maps of aluminum metal and oxide. a) Shows the simultaneously acquired ADF image and only the Al metal ($\sim 100$ nm thick) is visible. b), c) a layer of Al oxide was deposited on the left hand side. The EDS determined location is shown in atomic element map b) and the oxide map c). d) Shows the EDS chemical shift map in the two regions. e) The associated histogram of d). The measured shift between the Al metal and oxide is in good agreement with a separate dataset shown in the main text \fixme{(cite main text Fig. 4)} and result from the literature \cite{burkhalter_chemical_1972}}.
    \label{fig:ALMAP}
\end{figure}

\newpage
\section{Spectroscopy}
Which atoms are which? All atoms are about the same size ($\sim 1 \angstrom$), despite orders of magnitude variations in atomic number and mass across the periodic table. The last electron added to a single charged ion sees one positive nuclear charge. The similar Coulomb potentials make all atomic radii, to within factors of 1-3 \cite{slater_atomic_2004}, about the Bohr radius. Atom-atom bond distances are uniform to within factors of 1-3 \cite{brese_bond-valence_1991}. The lack of spatial diversity, finite sample thickness, and microscopic resolving power makes atomic and chemical bond identification by direct imaging difficult. Even using crystallography atoms must be assigned based on some prior information of chemical bonding or recovery of phase by direct methods \cite{sheldrick_phase_1990}. Identification by spectroscopy from their precise quantum mechanical energy levels is more robust. In addition, theses quantum mechanical measurements are decoupled from the microscope's resolving power, making them accessible to both macroscopic and microscopic imaging systems. 

How well we match the microscope's resolving power to spectroscopic spatial resolution is determined by electron source's brightness (commonly made from needle-shaped tungsten). Brightness is the electron current density $J_e$ cast through a solid angle $\Omega$, expressed mathematically as
\begin{equation}
    \beta=\frac{J_e}{\Omega}.
    \label{eq:bright}
\end{equation}
 The definition is similar (sans electron charge) to photon optics. More electrons are needed for spectroscopy than imaging. In scanning transmission electron microscopy (STEM), higher brightness sources accommodate more current in a smaller probe \cite{krivanek_atomic-resolution_2011} such that atomic resolution EELS \cite{gloter_atomically_2017} and EDS \cite{zaluzec_x-ray_2023} are now possible. Identifying atoms by their spectroscopic signature, and observing details such as chemical shifts, can tell how they are bonded.

Chemical shifts can be subtle and complex \cite{matensson_origin_1995,muller_why_1999}. A decrease in the valence electron density can decrease the Fermi energy, producing a negative chemical shift (smaller magnitude binding energy) \cite{lindgren_chemical_2004}. In contrast, opening a band gap can increase the lowest unoccupied level relative to the core state, creating an increased chemical shift (larger magnitude binding energy). In the latter case when a metal and oxide are brought together, electrons in the valence band move from the metal to the oxide \cite{muller_mapping_1993}, increasing the charge density in the oxide and causing the outer electrons to be less effectively screened, which in turn pulls all electrons closer to the nucleus and thereby increases the binding energy. There are other smaller effects such as the core-hole being effectively repulsed by the screened nuclear potential leading to a small negative chemical shift to lower binding energies, the relaxation case \cite{egerton_electron_2011}, and others. Transitions between different atomic orbitals are further complicated because the two energy levels may shift in different directions. When looking at changes in transitions rather than binding energies, one must be aware that the sign of the chemical shift can change because the orbitals being examined in a transition can move independently of each other.

\newpage
\section{Resolution}
Each successive source development, (the tungsten filament, LaB$_6$, Schottky field emission, the cold field emission, etc.), has produced roughly an order of magnitude increase in source brightness. The coherent current \cite{krivanek_atomic-resolution_2011} determines the amount of current that can be put into an $\angstrom$ sized probe. Additional current acts to broaden the focused electron probe. These diverge from the ideal wavelength limited resolution. The additional optical girth from the decoherence approximately adds in quadrature to the probe defined by the condenser aperture. Quantitatively we can express this as,
\begin{equation}
    d_{probe}=\sqrt{\left(\frac{0.61 \lambda}{\alpha}\right)^2  + \left( \sqrt{\frac{4I_e}{\pi\alpha^2 \beta}} \right)^2},
    \label{eq:btotal}
\end{equation}
where the probe size is $d_{probe}$. The diffraction of wavelength $\lambda$ from the limiting aperture outer angle $\alpha$ is combined quadratically with electron current $I_e$ and brightness $\beta$. As the brightness goes to infinity the probe is diffraction limited. In addition, more current put into our probe increases its size and thereby makes high resolution spectroscopy difficult. It is common to exchange resolution to access higher beam currents. For example, a 80 kV and a 10 mrad probe is roughly sized at 2.6 $\angstrom$, and if you include 500 pA and $\beta \approx 6\times10^{12}$ A/m$^2$/Sr (parameters used in the tungsten measurements) it swells to 11 $\angstrom$. The latter is far too large to resolve a crystalline lattice.

The probe is also broadened by inelastic collisions as it travels through the thickness of the sample \cite{egerton_limits_2007}. Electron beam broadening can increase the analytical x-ray volume and lead to a loss in spatial resolution \cite{hugenschmidt_electron_2019}. We can quantify lateral broadening $\Delta r$ as,
\begin{equation}
    \Delta r = C \frac{Z}{KE} n_a^{1/2}t^{3/2},
    \label{eq:thick}
\end{equation}
where the prefactor $C$ is 0.0045 keV nm, $Z$ is the atomic number, $n_a$ is atomic number density in 1/nm$^3$, $KE$ is the kinetic energy of the electron in (in keV), and $t$ is the thickness in nm \cite{hugenschmidt_electron_2019}. Tungsten (63 atoms/nm$^3$) has an atomic number density similar to aluminum (50 atoms/nm$^3$), but a $\approx$ 3.4 times higher atomic number. Multiple scattering produces a lateral spread of 2.8 nm (W) and 0.8 nm (Ti) through 10 nm of material at an accelerating voltage of 30 kV. This inelastic energy loss produces a large background in EELS that is not optimal for core-loss identification and chemical analysis at lower accelerating voltages. The flip side is that the energy loss is contributing to greater x-ray generation. For similar tungsten settings and 80 kV we find a thickness broadening of 1.1 nm, in total (combined with Equation \ref{eq:btotal}) the width becomes 1.5 nm. 

\newpage
\section{Supplementary EELS Information}
\begin{figure}[!htb]
\vspace{\SIspace in} 
    \includegraphics[width=\SIscale\linewidth]{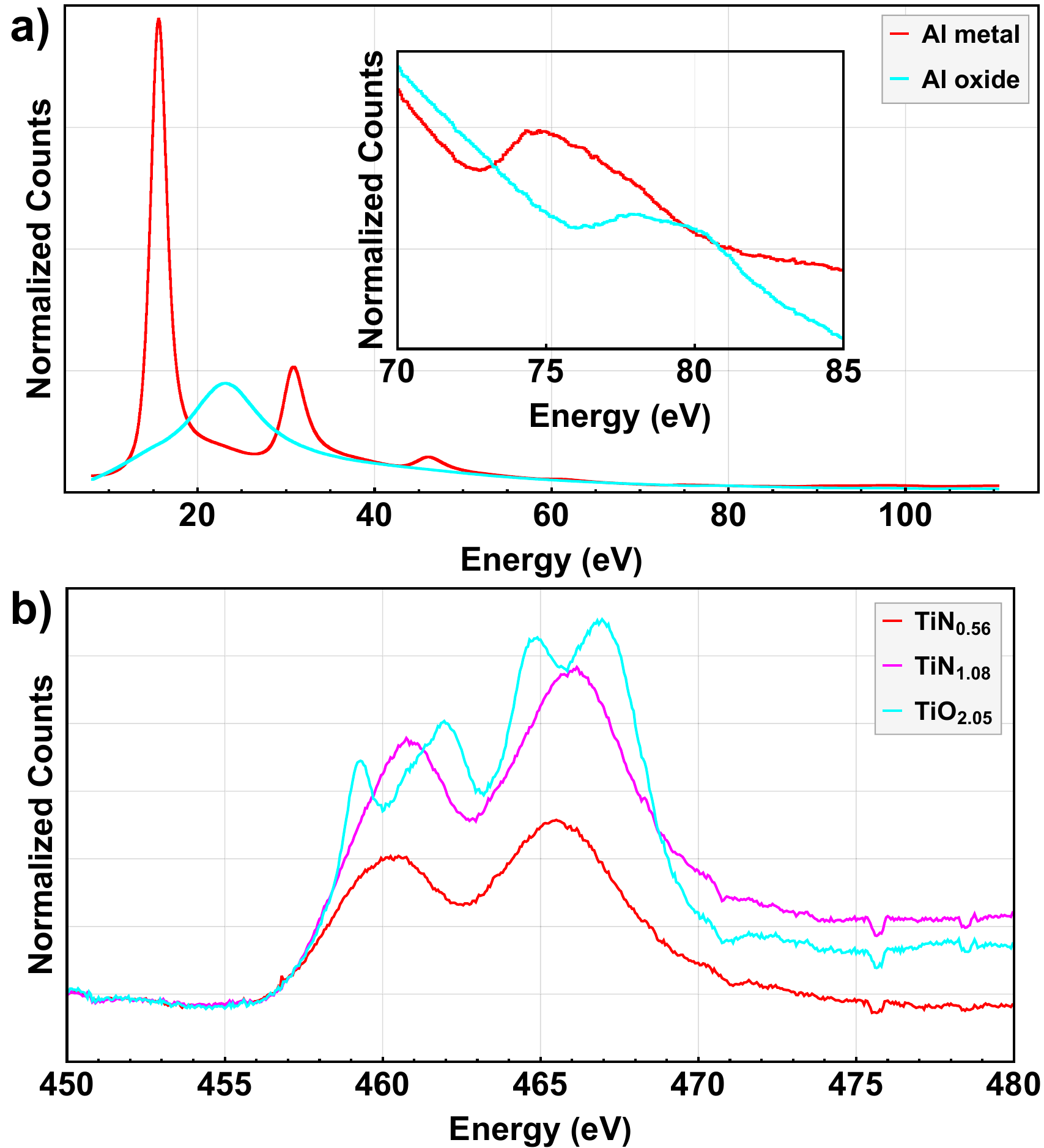}
    \captionsetup{width=\textwidth}
    \caption{a) The EELS spectrum from Al metal and oxide. The chemical shift is clearly visible in the inset plot. This dataset is summed from an ROIs in around the pure Al metal and oxide regions of data shown in main text Figure 4. b) EELS spectra of TiN$_{0.56}$, TiN$_{1.08}$, and TiO$_{2.05}$. Here the chemical shifts are not clearly distinguishable. Transitions in d-band orbitals split the fine structure peaks. The change in the binding energy are difficult to determine. Ti oxide as compared to the nitride \cite{dahle_silicon_2013, mizoguchi_first-principles_2009}. The front edge cannot be clearly delineated for a clean chemical shift determination, but the fine structure can a be used for fingerprinting at least with the oxide.}
    \label{fig:edssupE}
\end{figure}

\newpage
\begin{figure}[!htb]
\vspace{\SIspace in}   
    \includegraphics[width=\SIscale\linewidth]{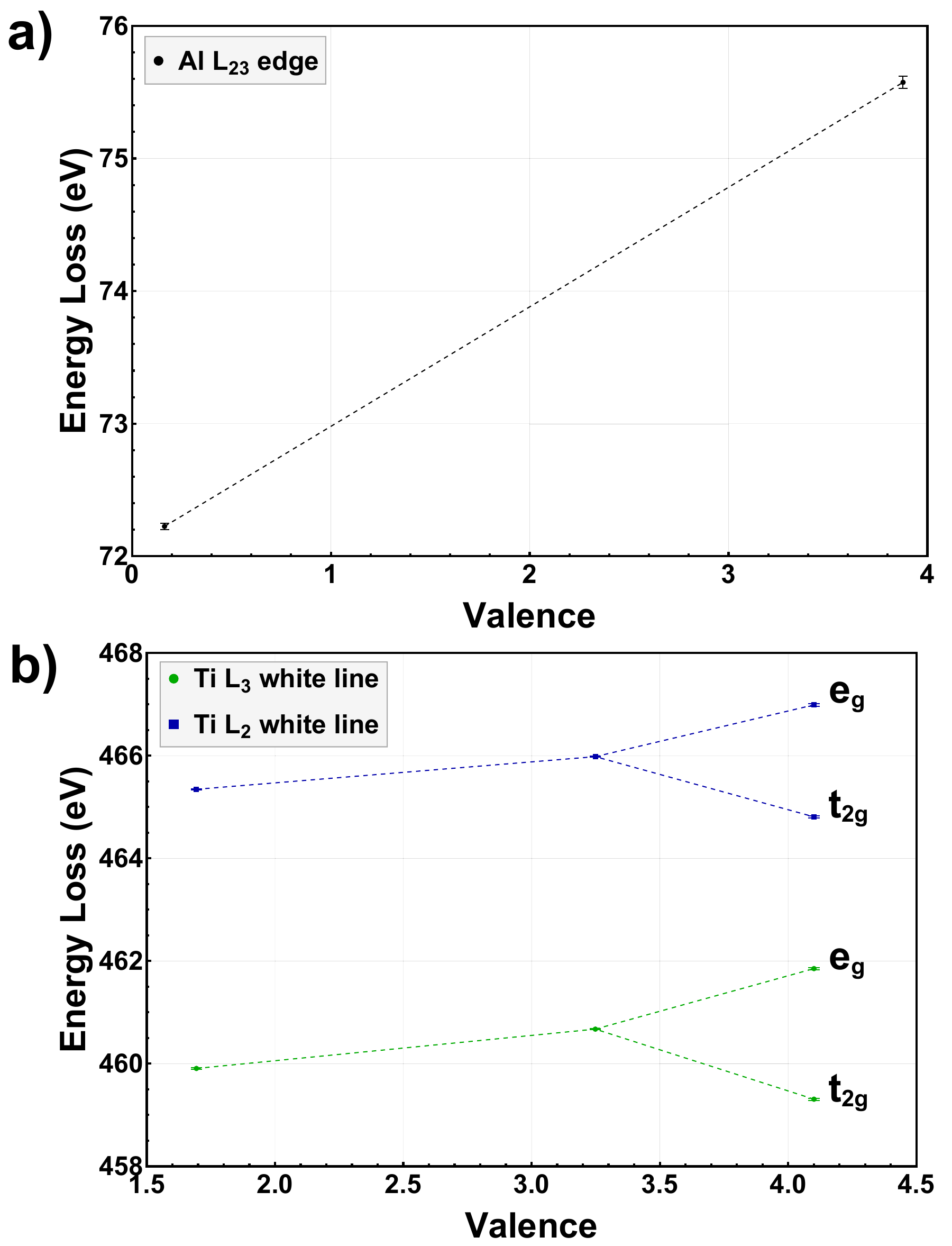}
    \captionsetup{width=\textwidth}
    \caption{EELS chemical shift as a function of the valence. The valence is determined by EDS in the main text. a) Aluminum $L_{23}$ edge energy (defined as the energy corresponds to 10\% level of the rising edge). b) Titanium $L_2$ and $L_3$ white line energies. Crystal field splitting in 3d orbitals is observed at a high oxidation state.}
    \label{fig:EELSCHEM}
\end{figure}

\newpage
\section{Tables of X-ray Parameters}

\begin{table}[!htb]
    \begin{tabular}{ |c |c |c |c |c|  }
        \hline
        Transition & Initial Binding Energy [eV] \cite{bearden_reevaluation_1967} & Final Binding Energy [eV] \cite{bearden_reevaluation_1967}& $\Delta$ Binding Energy [eV] \cite{bearden_reevaluation_1967} &  Measured X-ray [eV] \cite{bearden_x-ray_1967}\\ [0.5ex] 
        \hline\hline
        Al K$_{\alpha_1}$ & 73.1$\pm$0.5 & 1559.6$\pm$0.4 & 1486.5 & 1486.70\\
        \hline
        Al K$_{\alpha_2}$ & 73.1$\pm$0.5 & 1559.6$\pm$0.4 & 1486.5 & 1486.27 \\
        \hline
        Ti K$_{\alpha_1}$ & 455.5$\pm$0.4 & 4966.4$\pm$0.4 & 4510.9 & 4510.84 \\
        \hline
        Ti K$_{\alpha_2}$ & 461.5$\pm$0.4 & 4966$\pm$0.4 & 4504.9 & 4504.86 \\
        \hline
        Ti K$_{\beta_1}$  & 34.6$\pm$0.4  & 4966.4$\pm$0.4 & 4931.8 & 4931.81 \\
        \hline
        W M$_{\alpha_1}$  & 33.6$\pm$0.4  & 1809.2$\pm$0.3 & 1775.6 & 1775.4 \\
        \hline
        W M$_{\alpha_2}$  & 36.5$\pm$0.4 & 1809.2$\pm$0.3 & 1772.7 & 1773.1 \\
        \hline
        W M$_{\beta}$     & 36.5$\pm$0.4 & 1871.6$\pm$0.3 & 1835.1 & 1834.9 \\
        \hline
        W L$_{\alpha_1}$  & 1809.2$\pm$0.3 & 10206.8$\pm$0.3 & 8397.6 & 8397.6 \\
        \hline
        W L$_{\alpha_2}$  & 1871.6$\pm$0.3 & 10206.8$\pm$0.3 & 8335.2 & 8335.2 \\ 
        \hline
        W M$_{\zeta_1}$ &  425.3$\pm$0.5 & 1809.2$\pm$0.3  & 1383.9 &  1383.5 \\
        \hline
        W M$_{\zeta_2}$ & 491.6$\pm$0.4 & 1871.6$\pm$0.3 & 1380 & 1378.7 \\
        \hline
     \end{tabular}
    \caption{Table comparing the EELS levels (binding energies) to EDS transitions (binding energy changes). The binding energy changes and measured x-ray values are in good agreement to four digits, but very in the remaining precision by amounts sizable compared to EDS chemical shifts. This variation explains why existing measurements of binding energy difference cannot be used to predict the x-ray chemical shifts.}
    \label{fig:COMPARE}
\end{table}

\newpage

\bibliographystyle{apsrev4-1}
\bibliography{20240626_eds_cs_bibtex_unico_wo_mod}